\documentclass[12pt]{article}
\usepackage{color}
\usepackage{graphicx}
\pdfoutput=1

\input xy
\xyoption{all}
\xyoption{web}

\newlength{\abstractwidth}

\renewcommand{\thefootnote}{\fnsymbol{footnote}}
\renewcommand{\thanks}[1]{\footnote{#1}}
\newcommand{\starttext}{
\setcounter{footnote}{0}
\renewcommand{\thefootnote}{\arabic{footnote}}}

\newcommand{\bea}{\begin{eqnarray}}
\newcommand{\eea}{\end{eqnarray}}
\newcommand{\ee}{\end{equation}}
\newcommand{\be}{\begin{equation}}

\newcommand{\ea}{\end{array}}
\newcommand{\bac}{\begin{array}{c}}
\newcommand{\bacc}{\begin{array}{cc}}
\newcommand{\barcl}{\begin{array}{r@{}c@{}l}}
\newcommand{\brcl}{\begin{array}{rcl}}
\newcommand{\bdm}{\begin{displaymath}}
\newcommand{\edm}{\end{displaymath}}

\newcommand{\half}{\frac{1}{2}}

\def\cC{{\cal C}}

\def\cI{{\cal I}}
\def\cJ{{\cal J}}
\def\cK{{\cal K}}
\def\cL{{\cal L}}
\def\cM{{\cal M}}
\def\cN{{\cal N}}
\def\cO{{\cal O}}

\def\bC{{\bf C}}

\def\bR{{\bf R}}
\def\bZ{{\bf Z}}

\def\Re{{\rm Re}}
\def\Im{{\rm Im}}

\def\half{ {1\over 2}}
\def\p{\partial}

\def\a{\alpha}
\def\b{\beta}
\def\tet{\vartheta}
\def\ep{\varepsilon}

\def\g{\gamma}
\def\l{\lambda}

\def\o{\omega}

\def\f{\varphi}

\def\g{\gamma}

\def\np{{n '}}
\def\pp{{p '}}
\def\qp{{q '}}

\def\no{\nonumber}
\def\sm{\smallskip}

\begin{document}
\starttext
\setcounter{footnote}{0}

\begin{flushright}
IGC-12/5-1
\end{flushright}

\bigskip

\begin{center}

{\Large \bf Holographic duals of  Boundary CFTs}\footnote{This work is supported in part by NSF grants
PHY-08-55356 and PHY-07-57702.} 

\vskip 0.6in

{ \bf Marco Chiodaroli$^a$, Eric D'Hoker$^b$,  Michael Gutperle$^b$}

\vskip .2in

 ${}^a$ {\sl Institute for Gravitation and the Cosmos,}\\
{\sl The Pennsylvania State University, University Park, PA 16802, USA} \\
{\tt \small mchiodar@gravity.psu.edu}

\vskip 0.2in

 ${}^b$ { \sl Department of Physics and Astronomy }\\
{\sl University of California, Los Angeles, CA 90095, USA}\\
{\tt \small dhoker@physics.ucla.edu,  gutperle@physics.ucla.edu; }

\end{center}

\vskip 0.2in

\begin{abstract}

New families of regular half-BPS solutions to 6-dimensional Type 4b supergravity with $m$ tensor multiplets 
are constructed exactly. Their space-time consists of $AdS_2 \times S^2$ warped over a Riemann surface 
with an arbitrary number of boundary components, and arbitrary genus. The solutions have an arbitrary number 
of asymptotic $AdS_3 \times S^3$ regions. In addition to strictly single-valued 
solutions to the supergravity equations whose scalars live in the coset $SO(5,m)/SO(5)\times SO(m)$, we 
also construct stringy solutions whose scalar fields are single-valued up to transformations under 
the $U$-duality group $SO(5,m;\bZ)$, and live in the coset $SO(5,m;\bZ)\backslash SO(5,m)/SO(5)\times SO(m)$.
We argue that these Type 4b solutions are holographically dual to general classes of interface and boundary 
CFTs arising at the juncture of the end-points of 1+1-dimensional bulk CFTs. We evaluate their corresponding 
holographic entanglement and boundary entropy, and discuss their brane interpretation. We conjecture that 
the solutions for which $\Sigma$ has handles and multiple boundaries correspond to the near-horizon limit 
of half-BPS webs of dyonic strings and three-branes.

\end{abstract}

\newpage


\baselineskip=17pt
\setcounter{equation}{0}
\setcounter{footnote}{0}

\newpage

\section{Introduction and summary of results}
\setcounter{equation}{0}
\label{sec1}

The AdS/CFT correspondence relates string theory or M-theory on a space-time of 
the form $AdS_{d+1} \times M $ to a $d$-dimensional conformal field theory
\cite{Maldacena:1997re,Gubser:1998bc,Witten:1998qj}. Generally,
this CFT may be obtained directly from a system of branes in string theory or M-theory 
in the near-horizon limit where gravitational degrees of freedom decouple. The canonical
example is provided by the duality between Type IIB string theory on $AdS_5\times S^5$ 
and 4-dimensional $\cN=4$ super Yang-Mills theory obtained from the near-horizon limit 
of a stack of parallel  D3-branes.

\sm

The AdS/CFT correspondence may be applied with equal success to space-times 
which are only {\sl asymptotically} of the form $AdS_{d+1} \times M$, and their dual
field theories which are conformal only in the UV limit. One example includes field theories
with non-trivial renormalization group flow obtained by deforming a CFT by a relevant operator.
In a second example, a CFT  at finite temperature is dual to a gravity theory with a black
hole or black brane horizon. Field theories with finite charge density, spontaneously 
broken symmetries, and confinement may all be realized by different types of gravitational solutions.  
In all of the cases given above, scale invariance is broken by the introduction of dimensionful couplings. 

\sm

The field theories of interest to the present paper are characterized by the presence of an interface or 
a boundary.  An {\sl interface CFT} is obtained by gluing together several bulk CFTs
at a common interface. A {\sl boundary CFT} is obtained when a single bulk CFT ends on a boundary. 
When the interface or the boundary is flat, scale invariance may be preserved.
The classification and construction of such interface and boundary 
CFTs constitutes a key problem in the study of CFTs which enjoys several physical applications 
(see e.g. \cite{Cardy:1989ir} for a discussion of the two-dimensional case). While space-translations
transverse to the interface or boundary are no longer symmetries, the Poincar\'e transformations 
along the interface or boundary will be preserved. With a supersymmetric 
Yang-Mills theory in the bulk, various degrees of supersymmetry may be preserved by the 
interface or boundary, leading to theories with various superconformal symmetry invariances.
A classification for the $\cN=4$ case in 4 dimensions can be found in \cite{D'Hoker:2006uv,Gaiotto:2008sa}.

\sm

On the gravity side, there are two complementary constructions of superconformal  interface 
and boundary CFTs whose holographic duals are asymptotically $AdS_5 \times S^5$:\footnote{Recently, 
an exact identification between these two constructions was obtained  in 
\cite{Assel:2011xz,Aharony:2011yc,Benichou:2011aa}.}
\begin{enumerate}
 \item  Following \cite{Hanany:1996ie,Gaiotto:2008sd,Gaiotto:2008ak}, one considers intersecting 
brane configurations in flat space-time in which D3-branes can end on D5- and NS5-branes.
In the near-horizon limit  the world-volume theory on the D3-branes flows to a 4-dimensional 
interface or boundary CFT.  Various half-BPS brane configurations then provide a  natural 
classification of the boundary conditions which preserve a $OSp(4|4)$ superconformal symmetry 
\cite{D'Hoker:2006uv,Gaiotto:2008sa}. 
\item Fully back-reacted Type IIB supergravity solutions dual to an interface CFT can be constructed by 
using a Janus Ansatz  \cite{Bak:2003jk}, in which space-time is a fibration of $AdS_4\times S^2\times S^2$  
over a two-dimensional Riemann surface $\Sigma$. In \cite{D'Hoker:2007xy,D'Hoker:2007xz}\footnote{For half-BPS solutions in M-theory see \cite{D'Hoker:2008wc,D'Hoker:2008qm,D'Hoker:2009my}. For 
related  work  by other authors  see \cite{Clark:2005te,Gomis:2006cu,Lunin:2006xr,Lunin:2007ab,Yamaguchi:2006te,Gomis:2006sb,Kumar:2002wc,Kumar:2003xi,Lunin:2008tf}.} the corresponding half-BPS 
solutions with $OSp(4|4)$ symmetry were obtained exactly in terms of harmonic functions on $\Sigma$. 
A classification of all such half-BPS solutions  in terms of superalgebras was given in \cite{D'Hoker:2008ix}.
\end{enumerate}

\subsection{Holographic duals to Interface and Boundary CFTs}

A first goal of the present paper is to classify and construct the most general half-BPS solutions 
which are locally asymptotic to $AdS_3 \times S^3$, and holographically dual to an interface or 
a boundary CFT in 2 space-time dimensions. A second goal is to derive physical quantities of the dual CFT, 
such as the holographic interface or boundary entropy. A third goal is to exhibit the precise map 
between these holographic solutions and intersecting brane configurations in flat space-time, in analogy 
with the results described above for the $AdS_5\times S^5$ case. In this paper, we shall achieve 
the first two goals, and make progress towards the third.

\sm

The problem of constructing half-BPS solutions dual to interface and boundary CFTs in two dimensions
was attacked in \cite{Chiodaroli:2009yw,Chiodaroli:2009xh} directly in the 
language of Type IIB supergravity. There, space-time was of the form $AdS_2 \times S^2 \times K3$ 
warped over a Riemann surface $\Sigma$ with an arbitrary number of boundary components,
but no handles. The solutions obtained are quarter-BPS from the point of view of 10-dimensional 
Type IIB supergravity, and thus invariant under 8 residual supersymmetries. In these solutions, 
however, only the volume form of the $K3$ manifold played a role, while the 
other homology generators of the $K3$, and their associated moduli, were turned off. Moreover, the 
supergravity fields, including the axion, were strictly single-valued, thereby excluding interesting
quantum solutions.

\sm

The problem was also attacked in \cite{Chiodaroli:2011nr} in the language of the 6-dimensional
Type~4b supergravity, a theory which arises as the dimensional reduction of Type IIB on $K3$. 
From the point of view of 6-dimensional Type 4b theory, the solutions are half-BPS and are
again invariant under 8 residual supersymmetries.
Type~4b supergravity has 5 self-dual and $21$ anti-self-dual  3-form fields, as well as 105 scalars 
living on the $SO(5,21)/SO(5)\times SO(21)$ coset \cite{Romans:1986er}. This formulation 
has the distinct advantage of exhibiting the moduli of all $K3$ deformations on the same footing
as the dilaton-axion scalars of Type IIB, and thus making the full U-duality group $SO(5,21;\bZ)$ manifest.

\sm

Type 4b supergravity lends itself well to identifying self-dual string solutions which carry the 
self-dual 3-form charge. Six-dimensional self-dual strings arise effectively from certain bound 
states of branes in 10-dimensional  Type IIB string theory.  The D1/D5 system with 4 worldvolume dimensions 
wrapped on $K3$ provides one example of such a self-dual string (the others can be obtained by 
the action of the U-duality group).  In the near-horizon limit a self-dual string produces a supersymmetric 
$AdS_3\times S^3$ vacuum, invariant under the global superalgebra $PSU(1,1|2)\times PSU(1,1|2)$.

\sm

Much like $(p,q)$ strings of Type IIB in 10-dimensional flat Minkowski space-time 
\cite{Schwarz:1996bh,Dasgupta:1997pu,Sen:1997xi},  junctions  where $N$  self-dual strings 
of Type 4b supergravity  come together  can be constructed in 6-dimensional Minkowski space-time.
As discussed in \cite{Chiodaroli:2010mv} the supersymmetry algebra of 6-dimensional 
Type~4b supergravity contains a central charge $Z_\mu^i$ which is a Minkowski vector 
as well as a vector with respect of the $SO(5)$ R-symmetry. The central charge  depends on the self-dual 
string charge and the spatial orientation of the self-dual string. 
 Networks of self-dual strings are ${1\over4}$-BPS in flat space-time provided 
they are planar and there is an alignment of the the spatial and $SO(5)$ directions of the central charge.

\sm

Apart from the self-dual strings, other supersymmetric branes exist in the  6-dimensional 
Minkowski vacuum. In particular, three-branes can be obtained from  D3-branes, 5-branes 
wrapping 2-cycles of the $K3$, and D7-branes wrapping the whole $K3$.  Since these branes  
are co-dimension two from the point of view of the 6 non-compact dimensions they  induce 
non-trivial monodromy on the axionic scalars which enjoy a shift symmetry  in the 
$SO(5,21)/SO(5)\times SO(21)$ coset.  A novel feature of networks with three-branes is that  self-dual 
strings can end  on them.  Charge conservation holds as the charge of the string is  absorbed 
on the world-volume of the three-brane \cite{Callan:1997kz}.

\subsection{Novelties of the present solutions}

In the present paper  we shall complete the construction, which was initiated in 
\cite{Chiodaroli:2009yw,Chiodaroli:2009xh,Chiodaroli:2011nr}, of the general half-BPS 
holographic solutions in 6-dimensional Type 4b supergravity, and thus in 10-dimensional Type IIB,  
which are dual to interface and boundary 
CFTs in two dimensions\footnote{Type 4b supergravity with $m = 21$ 
can be obtained from the Kaluza-Klein compactification of Type IIB supergravity on K3. While it is
unclear whether in general this truncation is consistent or not, solutions of the theory in six dimensions
should uplift to 10-dimensional solutions in the limit in which the volume of the K3 is small.}. The construction follows the Janus Ansatz with space-time geometry
$AdS_2\times S^2$ warped over a Riemann surface $\Sigma$ with boundary.  
The novel results of the present solutions are as follows:
\begin{enumerate}
\itemsep=0in
\item Building on the general local solutions of \cite{Chiodaroli:2011nr}, the scalar fields in the 
present solutions take values in the largest sub-manifold of the scalar coset 
$SO(5,m)/SO(5)\times SO(m)$ allowed by the half-BPS condition, namely $SO(2,m)/SO(2)\times SO(m)$.
By contrast, the solutions of \cite{Chiodaroli:2009yw,Chiodaroli:2009xh} were restricted to 
$SO(2,2)/SO(2)\times SO(2)$.
\item The present solutions allow for monodromy of the axionic scalars by elements of the U-duality group
$SO(5,m;\bZ)$. This gives rise to stringy solutions in which the scalar fields live on a sub-manifold
of the coset $SO(5,m;\bZ)\backslash SO(5,m)/SO(5)\times SO(m)$. 
By contrast, the scalar 
fields in the solutions of \cite{Chiodaroli:2009yw,Chiodaroli:2009xh,Chiodaroli:2011nr} 
were single-valued only.
\item The present solutions include surfaces $\Sigma$ with an arbitrary number of handles,
producing an associated increase in the number of moduli. We conjecture that these solutions 
provide holographic duals to half-BPS networks of self-dual strings and three-branes in Type 4b. 
By contrast, the solutions of \cite{Chiodaroli:2009yw,Chiodaroli:2009xh,Chiodaroli:2011nr}
excluded handles. 
\end{enumerate}

\subsection{Overview of solutions dual to  interface and boundary CFTs}

In the sequel of this introduction,  we shall present an overview of  the properties of all 
solutions, old and new,  in more detail. The first ingredient in the solutions is a real harmonic function $H$
which is positive inside $\Sigma$ and vanishes on its boundary $\p \Sigma$, except at a finite 
number $N$ of simple poles on $\p \Sigma$. The region near each pole corresponds to a locally 
asymptotic $AdS_3\times S^3$ which is dual to a ``bulk" CFT living on a half-space $\bR^+\times \bR$. 
The second ingredient is a set of $m$ meromorphic functions $L^A$, which are holomorphic 
inside $\Sigma$, are required to be real on $\p \Sigma$, and have a finite number of simple poles 
on $\p \Sigma$.

\begin{figure}[htb]
\begin{center}
\includegraphics[scale=.5]{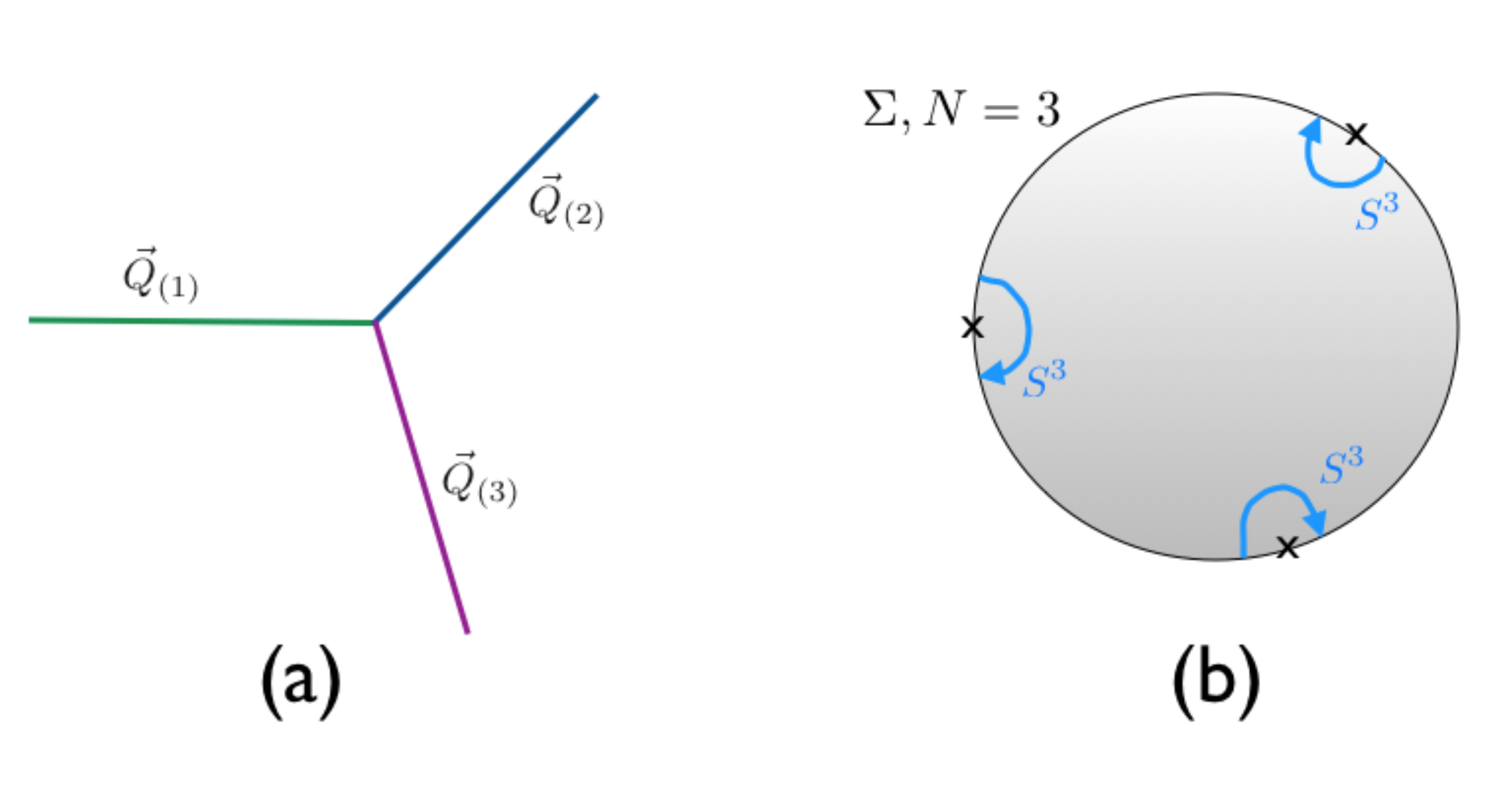} 
\caption{\small
(a)  A junction of three self-dual strings  in 6-dimensional flat space-time. 
(b) The corresponding Riemann surface $\Sigma$ for the dual holographic solution with $N=3$ poles. 
The blue cycles indicate three spheres on which the 3-form charges are supported.}
\label{draft-fig1}
\end{center}
\end{figure}


The simplest solutions are based on  a Riemann surface $\Sigma$  with a single 
boundary component and no handles, namely  the disk. They were constructed in \cite{Chiodaroli:2011nr}, 
and will be briefly reviewed in Section  \ref{sec2}. These solutions are holographic realizations of 
string junctions, where~$N$ two-dimensional CFTs are glued together along a one-dimensional interface. 
The example of a three-string junction and its corresponding  surface $\Sigma$ is depicted in Figure \ref{draft-fig1}. 
These solutions can be obtained by taking a decoupling limit of   ${1\over4 }$-BPS junction of $N$ 
self-dual strings  in flat space-time. One argument for this identification is  the exact match 
between the parameters of the supergravity solution and the physical quantities characterizing the 
string junction.  Namely, the parameters of the solution can be identified with  the 3-form charges 
 on  the asymptotic $S^3$, and the values of the scalars   in the asymptotic region 
which are not fixed by the attractor mechanism.

\sm
 
The regular solutions constructed in \cite{Chiodaroli:2011nr} are dual to interface CFTs for $N=2$, 
and to the junction of $N$ CFTs for $N>2$.  The dual to a boundary CFT should have $N=1$.
Recently there has been a renewed interest in the holographic 
description of boundary CFTs \cite{Karch:2000gx,Takayanagi:2011zk,Fujita:2011fp,Nozaki:2012qd,Kwon:2012tp,Setare:2011ey,Alishahiha:2011rg}. 
No regular solutions with only one boundary were found in \cite{Chiodaroli:2011nr}, but 
singular half-BPS solutions with $N=1$ were constructed in \cite{Chiodaroli:2011fn} 
as limits of regular interface solutions. The new type of singularities  were referred to as the  
$AdS_2$-cap and $AdS_2$-funnel in \cite{Chiodaroli:2011fn}. 

\sm

\begin{figure}[h]
\begin{center}
\includegraphics[scale=.52]{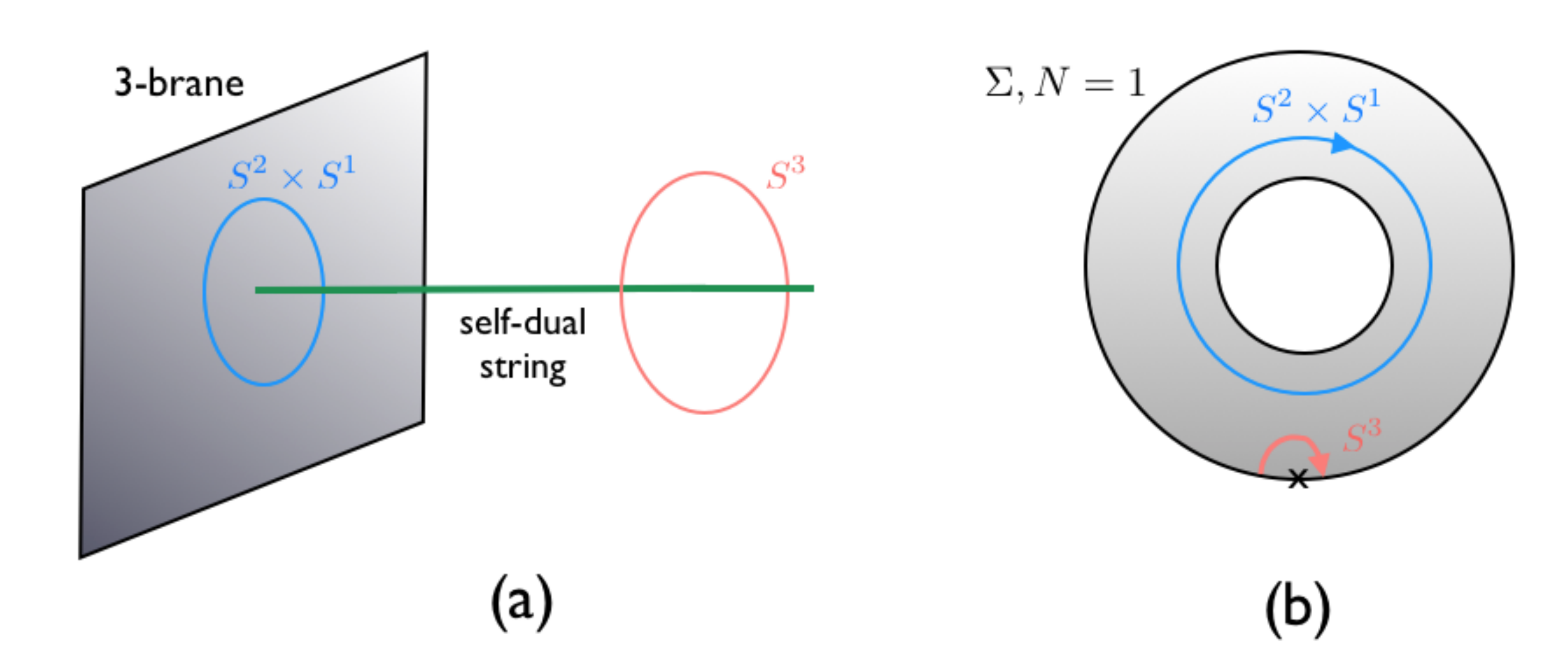} 
\caption{\small
(a)  A self-dual string ending on a three-brane in six dimensions. 
The charge of the string is absorbed on the three-brane.
(b) Riemann surface $\Sigma$ for the dual BCFT  solution with one  pole. 
Near the pole a $S^3$ supports the charge, there is a second non-contractible 
$S^1\times S^2$ cycle.  }
\label{draft-fig2}
\end{center}
\end{figure}

Perhaps the most important new result of the present paper is the construction of an infinite class of 
{\sl regular half-BPS solutions} which are holographically dual to boundary CFTs. 
The simplest non-trivial case arises when  $\Sigma$ is an annulus,
with a single asymptotic $AdS_3 \times S^3$ region. The configuration is illustrated 
in Figure \ref{draft-fig2}, and will be solved for in detail in Section~\ref{sec5}. 
The solution has non-trivial monodromy of the axionic scalars, and thus carries three-brane charge. 
Thus,  we propose that this annulus solution is the holographic dual to a BCFT 
which is obtained as a decoupling limit of a self-dual string ending on a D3-brane. 

The annulus solution has two new features compared to the solutions on the disk.  First, apart from the 
$S^3$-cycle associated with the asymptotic $AdS_3 \times S^3$ region, there is a $S^2\times S^1$-cycle  
which appears because the annulus has a non-contractible  $S^1$-cycle. This cycle can carry 3-form 
charge and hence charge conservation can be obeyed for a regular solution even when only a single asymptotic 
$AdS_3\times S^3$ region is present.  Second, when transported along the non-contractible cycle of the annulus the 
scalars living in the $SO(5,21)/SO(5)\times SO(21)$ coset pick up a non-trivial Abelian monodromy. 
In supergravity, such a solution is not single-valued and therefore not regular. 
However, the stringy quantum moduli space of the theory is quotient of the $SO(5,21)/SO(5)\times SO(21)$ 
coset  by the discrete  U-duality group $SO(5,21;\bZ)$. If the monodromy identifies scalars by an Abelian 
subgroup of the U-duality group  (i.e. a shift of axionic scalars)  the resulting solution is regular 
in string theory. This monodromy indicates the presence of a three-brane in the 6-dimensional 
space-time\footnote{For a discussion of such branes in the probe approximation see e.g.   \cite{Bachas:2000fr,Raeymaekers:2006np}.}.

\sm

In Section \ref{sec6} we obtain the most general regular BPS solutions  by considering 
Riemann surfaces $\Sigma$ which have $\nu$ boundary components, $g$ handles, 
and $N$ asymptotic $AdS_3\times S^3 $ regions.   The construction involves the double 
$\hat \Sigma$ of  $\Sigma$, which is a compact Riemann surface of genus $\hat g = 2g +\nu-1$. 
The harmonic and holomorphic functions which parametrize the solution are constructed using 
the prime form and holomorphic differentials, objects which are familiar from multi-loop string 
perturbation theory \cite{D'Hoker:1988ta}. 
As for  the annulus (which is a special case of the general solution with $g=0, \nu=2$) 
the solution displays non-trivial monodromy around non-contractible cycles.  
The solution is a proper string theory solution provided the scalar monodromy around each 
cycle is a U-duality transformation.

\subsection{Conjecture on holographic duals to string networks}

For any surface $\Sigma$, a solution whose harmonic function $H$ has $N$ poles 
can be interpreted as a holographic dual  
of a junction of $N$ CFTs each one living on a half-space. The charges and the values of the  
non-attracted  scalars in each asymptotic region completely determine the CFTs which 
are glued together.  The remaining parameters of the solution (which 
exist for all $\Sigma$ but the disk) should determine the gluing/boundary conditions at 
the junction. Hence the holographic solutions provide  an infinite set of boundary and 
interface CFTs. It should be possible to calculate CFT junction correlation functions holographically, 
for any solution depending on its full set of free parameters, though technically this may be involved.

\begin{figure}[h]
\begin{center}
\includegraphics[scale=.52]{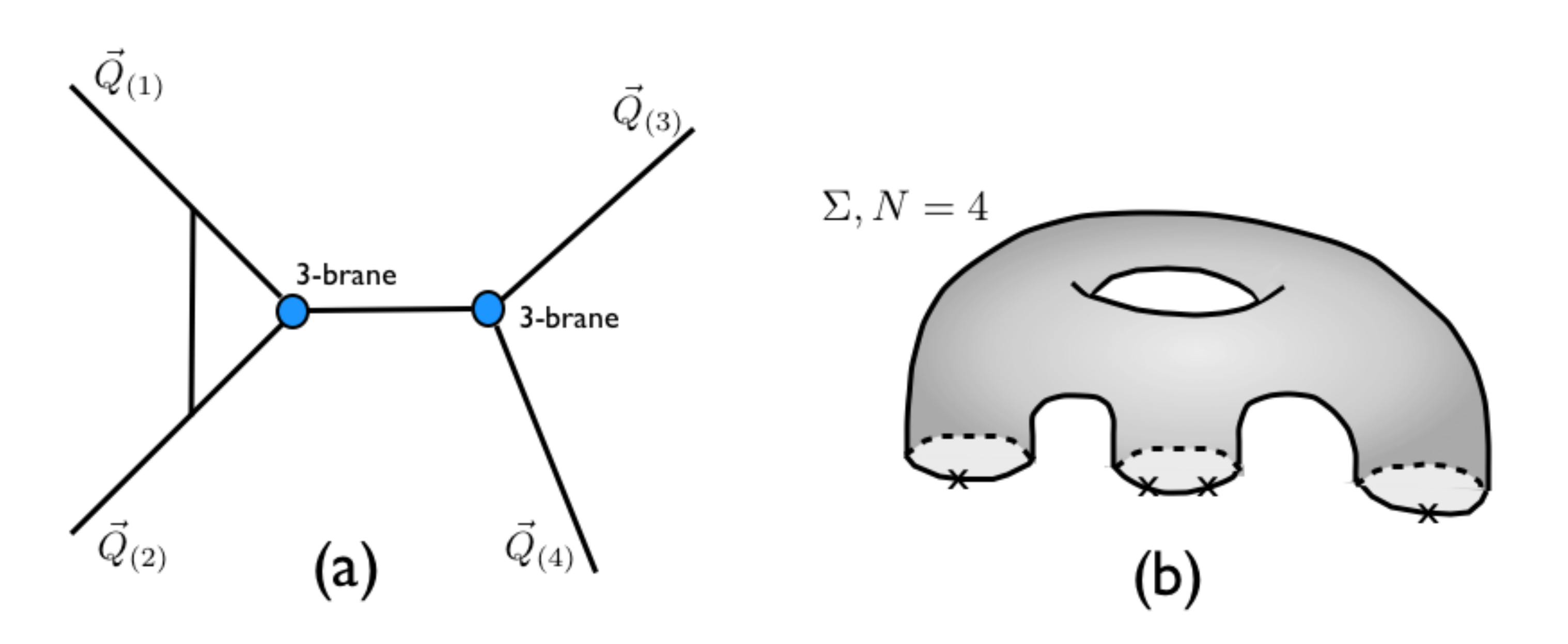} 
\caption{\small
(a)  A configuration of three-branes and self-dual string networks. 
In the IR the system flows to a interface CFT.
(b) A Riemann surface $\Sigma$ with $N=4$, $\nu=3$ and $g=1$.}
\label{draft-fig3}
\end{center}
\end{figure}

\vskip -0.1in

The new feature of solutions with multiple boundary components are the non-trivial scalar 
monodromy and, relatedly, the presence of three-branes.  For the annulus solution we provided 
some evidence that such a solution can be obtained from a decoupling/near-horizon limit 
of a self-dual string ending on a three-brane in 6-dimensional flat space-time.

\sm

We conjecture that the most  general solutions with arbitrary $g$ and $\nu$ 
can be obtained by taking a decoupling limit of a   network of  
self-dual string and   three-branes in flat space-time (see Figure \ref{draft-fig3}a for an example of such a configuration). 
This situation should be analogous to the Gaiotto-Witten construction of 4-dimensional boundary 
and interface CFT discussed in the beginning of the Introduction. We have so far not 
been able to obtain a precise  identification of the general supergravity solutions 
with a string and brane network. In particular the role of the handles is  not well 
understood. We leave these question for future investigation.

\sm

The remainder of this paper is organized as follows. In Section \ref{sec2}, we 
briefly review Type~4b supergravity, the construction of half-BPS solutions  by
warping $AdS_2 \times S^2$ over a Riemann surface $\Sigma$, and the solutions on the disk.
In Section \ref{sec3}, we construct the general regular solutions for the case where $\Sigma$ is an annulus,
and allow for solutions with monodromy in the U-duality group. In Section \ref{sec5}, 
the annulus solution for a single asymptotic $AdS_3 \times S^3$ is worked out in 
detail, the associated boundary entropy is evaluated, and its degeneration is obtained. Finally, in Section \ref{sec6},
the most general regular half-BPS solution is constructed in terms of a Riemann surface $\Sigma$
with arbitrary number of boundary components, arbitrary number of asymptotic $AdS_3 \times S^3$ 
regions, and arbitrary number of handles. We defer to the Appendix the lengthy calculation of the 
boundary entropy for the annulus.

\newpage

\section{Review of regular half-BPS solutions}
\setcounter{equation}{0}
\label{sec2}

In this section we shall briefly review the construction of half-BPS string-junction solutions in 
6-dimensional Type $4b$ supergravity carried out in \cite{Chiodaroli:2011nr}, which may be 
consulted for detailed derivations of the results.

\subsection{ Six dimensional Type $4b$ supergravity}

The theory was constructed in \cite{Romans:1986er}. We shall follow the notations and conventions 
of \cite{Chiodaroli:2011nr}.  The supersymmetry of the Type $4b$ theory is generated by two  
 symplectic-Majorana  spinors. Type 4b  contains a supergravity multiplet and $m$ tensor 
multiplets.  Anomaly cancellation restricts the value of $m$ to the case of either $m=5$ or 
$m=21$, corresponding respectively to compactification of Type IIB on $T^4$ or $K3$. 
For definiteness we will principally work with the $K3$ compactification and choose $m=21$ 
in the sequel.\footnote{Throughout, the index $A=(I,R)$ will label the fundamental 
representation of $SO(5,m)$, while the indices $(i,r)$ will label the fundamental representation 
of $SO(5) \times SO(m)$. Their ranges are given by, $I,i=1,\ldots , 5$ and $R,r=6, \ldots, m+5$ respectively.}
The supergravity multiplet contains the metric and the Rarita-Schwinger field as well as   five
rank-two anti-symmetric tensors $B^I$. Each tensor multiplet contains an anti-symmetric rank-two tensor 
$B^R$, and a quartet of Weyl fermions whose chirality is opposite to that of the gravitini.
Finally, the scalar fields  live in the real $SO(5,m)/(SO(5)\times SO(m))$ coset space and are
parametrized by a real frame field $V^{i,r}_{\;\;A}$. They may be represented by composite 
1-form fields $P^{ir}$, $Q^{ij}$ and $S^{rs}$ which are the block components of $V^{-1} dV$.
The 2-form potentials $B^A_{\mu \nu}$ give rise to field strength 3-forms,
$G^A = d B^A$. The associated $SO(5,m)$-covariant field strength 3-forms 
$H^i $ and $H^r$, which are respectively self-dual and anti-self-dual, obey, 
\bea
\label{2b2}
H^i  = V^i{}_A \, G^A & \hskip 1in & *H^i = + H^i  
\no \\
H^r  = V^r{}_A \, G^A && *H^r = - H^r 
\eea
The fields $H^i$ and $H^r$ obey Bianchi identities which follow from $dG^A=0$,
\bea
\label{2c3}
dH^i - Q^{ij} \wedge H^j - \sqrt{2} P^{ir} \wedge H^r & = & 0
\no \\
dH^r - S^{rs} \wedge H^s - \sqrt{2} P^{ir} \wedge H^i & = & 0
\eea
For the field strengths $H^i$ and $H^r$, the Bianchi identities and field equations 
are equivalent to one another in view of their duality properties. The Einstein equations are given by,
\be
\label{2d3}
R_{\mu\nu}- H^{i}_{\mu\rho\sigma}H^{i\; \rho\sigma}_{\nu}
- H^{r}_{\mu\rho\sigma}H^{r\; \rho\sigma}_{\nu}-2 P_{\mu}^{ ir} P_{\nu}^{ir}=0
\ee
The field equation for the scalars is given by,
\be
\label{2d4}
\nabla^{\mu} P_{\mu}^{ir} - (Q^{\mu})^{ij}   P_{\mu}^{jr} - (S^{\mu})^{rs} P_{\mu}^{is}  
-{\sqrt{2}\over 3} H^{i\; \mu\nu\rho}H^{r}_{\mu\nu\rho}=0
\ee
The fermionic fields $\psi _\mu ^\a$ and $\chi ^{r\a}$ of the Type 4b supergravity,
as well as the local supersymmetry spinor parameter $\ep^\a$, have definite chiralities,
and obey the symplectic Majorana conditions.
The BPS equations  are given by,\footnote{The Dirac matrices $\g^\mu$ with respect to 
coordinate indices are related to
the Dirac matrices $\g^M$ with respect to frame indices by  $\g^M = e^M {}_\mu \g^\mu$.}
\bea
\label{2e3}
0 &=&
\partial_{\mu} \ep ^\a
+ {1\over 4} \omega_{\mu}^{MN}\gamma_{MN} \ep ^\a
-{1\over 4} Q_{\mu}^{ij} (\Gamma^{ij})^{\alpha} {} _\b\, \ep^\beta 
-{1\over 4} H^{i}_{\mu\nu\rho} \gamma^{\nu\rho} \;
(\Gamma^{i})^{\alpha}_{\;\; \beta}\ep^{\beta}
\no\\
0 &=&
{1\over \sqrt{2}} \gamma^{\mu} P_{\mu}^{ir}
(\Gamma^{i})^{\alpha}_{\;\; \beta} \ep^{\beta}
+{1\over 12} \gamma^{\mu\nu\rho}H^{r}_{\mu\nu\rho}\ep^{\alpha}
\eea
The theory has a 6-dimensional Minkowski space-time vacuum where all anti-symmetric tensor 
fields are set to zero,
and which is invariant under 16 Poincar\'e supersymmetries. Self-dual BPS string solutions  
can be constructed which preserve 8 of the 16 Poincar\'e supersymmetries.   
The $AdS_3\times S^3$ vacuum  with 16 supersymmetries and 
$PSU(1,1|2)\times PSU(1,1|2) $ symmetry superalgebra emerges in the near-horizon 
limit.

\subsection{Regular half-BPS  solutions}\label{sec2b}

The geometry of a local half-BPS string-junction solution in six dimensions consists of an
$AdS_2 \times S^2 $ space warped over a two-dimensional Riemann surface $\Sigma$ 
with boundary \cite{Chiodaroli:2009yw,Chiodaroli:2011nr}. 
The metric and anti-symmetric tensor fields of the solution are, 
\bea
\label{ansatz}
ds^2 & = & f_1^2 ds^2 _{AdS_2} + f_2 ^2 ds^2 _{S^2} +  ds^2 _\Sigma 
\no \\
B^A & = & \Psi ^A \, \o_{AdS_2} + \Phi ^A \, \omega _{S^2}
\eea
Here, $ds^2 _{AdS_2}$ and $ds^2 _{S^2}$ are the invariant metrics respectively on the spaces 
$AdS_2$ and $S^2$ of unit radius, while $\o_{AdS_2}$ and $\omega _{S^2}$ are the corresponding 
volume forms. In local complex coordinates $(w,\bar w)$ on $\Sigma$, the metric  
$ds_\Sigma ^2 = 4 \rho ^2 |dw|^2$ is parametrized by a real function~$\rho$. 

\sm

The general local solution was constructed in \cite{Chiodaroli:2011nr} and 
is specified completely in terms of a real positive harmonic function 
$H$ and $m+2$ meromorphic functions $\lambda^A$ on $\Sigma$.  
Solving the BPS equations, Bianchi identities, and field equations, subject
to the reality of the fields, imposes the restriction 
$\lambda^3=\lambda^4=\lambda^5=0$, (up to $SO(5,m)$ 
transformations), as well as,\footnote{Throughout, we shall use the $SO(5,m)$-invariant
metric $\eta = {\rm diag}(I_5, -I_m)$ to raise and lower indices $K^A = \eta ^{AB}  K_B$ for 
vectors $K \in \bC ^{5,m}$, and to form their inner invariant product  $K \cdot L = \eta _{AB} K^A L^B$.}
\bea
\label{1b2}
\l \cdot \l & = & 2
\no \\
\bar \l \cdot \l & \geq & 2
\eea
with strict inequality of the second condition in the interior of $\Sigma$.
The solution of \cite{Chiodaroli:2011nr} provides explicit formulas 
for the metric factors,
\bea
\label{7c1}
f_1 ^4 = H^2 ~ 
{ \bar \lambda \cdot \lambda  +2 \over \bar \lambda \cdot  \lambda  -2} 
& \hskip 1in & 
\rho^4 = {|\partial _w H|^4 \over 16 H^2} 
(\bar \lambda \cdot  \lambda  +2)  ( \bar \lambda \cdot \lambda  -2)
\no \\
f_2 ^4 = H^2 ~ 
{ \bar \lambda \cdot  \lambda  -2 \over \bar \lambda \cdot \lambda  +2} 
&&
\eea
On the solutions, the scalar field $V$ takes values in an $SO(2,m)/( SO(2) \times SO(m))$ 
sub-manifold of the real coset space $SO(5,m)/( SO(5) \times SO(m))$, so that $V$ takes the form,  
\bea 
V = \left( \begin{array}{ccc} V^i {}_I & 0 & V^i {}_R \cr
0 & I_3 & 0 \cr
V^r {}_I & 0 & V^r {}_R \cr   \end{array} \right) 
\eea
where $i,I=1,2$, and $r,R=6, \ldots, m+5$. The block $I_3$ represents the identity in the 
indices $i,I=3,4,5$. The combinations $V^\pm_{I,R}=V^1_{I,R} \pm i V^2_{I,R} $ may 
be expressed in terms of the functions $H$ and $\lambda^A$, up to an overall phase 
which depends on the choice of gauge, by the relation, 
\bea
\label{7b9}
 V^+{}_A = X ( \bar \lambda _A -  |X|^2 \lambda _A) /( 1 - |X|^4)
\eea
and its complex conjugate $V^-_A$, 
where $A=1,2,6,7, \ldots, m+5$, and $|X|$ is given by $|X|^2 + |X|^{-2} = \l \cdot \bar \l $.
The remaining entries of $V$  contain no further degrees of freedom and  
may be derived from the defining property for the  $SO(5,m)$ frame,  
$V^{-1} = \eta V^t \eta$. On the solutions, the anti-symmetric tensor fields $B^A$ are subject 
to the restriction $B^3=B^4=B^5=0$, which parallels the restriction of the scalars. 
The solutions for the remaining components of the 
real-valued flux potential functions $\Phi ^A$ and $\Psi ^A$ are as follows,
\bea
\label{7c3}
\Phi ^A =  -   \sqrt{2} \, {  H \, \Re( \lambda^A) 
\over \bar \lambda \cdot \lambda +2 } + \tilde \Phi ^A 
& \hskip 0.4in &
\tilde \Phi^A = { 1 \over 2 \sqrt{2} } \int \p_w H \, \lambda^A   +{\rm c.c.} 
\no \\
\Psi ^A =  - \sqrt{2} \,   { H \, \Im (\lambda^A) 
\over \bar \lambda \cdot \lambda -2 } + \tilde \Psi ^A 
& \hskip 0.4in &
\tilde \Psi^A = { i \over 2 \sqrt{2} } \int  \p_w H \, \lambda^A  +{\rm c.c.} 
\eea
where again $A=1,2,6,7,\ldots, m+5$.

\sm
 
The metric and other fields will be regular throughout $\Sigma$, provided  
the harmonic function $H$ and the meromorphic functions $\l^A$ satisfy the 
following regularity requirements, 
\begin{description}
\item{~~~1.} In the interior of $\Sigma$ we have $H>0$ and $\bar \lambda \cdot \l >2$;
\item{~~~2.} On the boundary $\p \Sigma$ of $\Sigma$ we have $H=0$ and $\Im (\lambda ^A) =0$,
except at isolated points;
\item{~~~3.} The 1-forms $\lambda ^A \p_w H$ are holomorphic and nowhere 
vanishing in the interior 
of $\Sigma$, forcing the poles of $\lambda^A$ to coincide with the zeros of $\p_w H$;
\item{~~~4.}  The functions $\l^A$ are holomorphic near $\p \Sigma$, 
thereby allowing for poles in $\lambda ^A \p_w H$ on $\p \Sigma$ only at those points 
where $\p_w H$ has a pole.
\end{description}

In \cite{Chiodaroli:2011fn}, a mild relaxation of some of these regularity conditions 
was shown to lead to generalized solutions, which are still physically acceptable.
They include the $AdS_2$-cap and the $AdS_2$-funnel, which may be interpreted 
as fully back-reacted brane solutions and give simple, though singular, examples of 
holographic boundary CFTs.

\subsection{Light-cone variables}
\label{twothree}

A general explicit solution of the constraints (\ref{1b2}) and the regularity requirements $1-4$ of Section 
\ref{sec2b} may be obtained following \cite{Chiodaroli:2011nr} by parametrizing the 
functions $\lambda ^A$  in terms of \emph{light-cone variables} $L^A$, defined by,
\bea
\label{8a1}
\l^A = { \sqrt{2} \, L^A \over L^6} \hskip 0.4in & \hskip 0.8in & 
\l^+ =  {\sqrt{2} \over L^6} \left ( - L^1 L^1 + L^R L^S  \delta _{RS} \right )
\no \\ 
\l^\pm = { 1 \over \sqrt{2}} (\l^2 \pm  \l^6) && \l^- = { 1 \over \sqrt{2} L^6}
\eea
where the indices run over $A=1,7,8,\ldots, m+5$, and 
$R,S=6,7,8, \ldots, m+5$. This parametrization explicitly solves the first constraint of 
(\ref{1b2}). Multiplication of $L^6$ and $L^A$ by a common constant $\ell$ leaves the $\l^A$
unchanged, but transforms $\l^\pm$ by an $SO(5,m)$ {\sl boost}, $\l^\pm \to \ell ^{\pm 1} \l^\pm$. 
Since the $\lambda ^A$ 
are meromorphic, the functions $L^A$ must also be meromorphic. Moreover, the 
harmonic functions $h^A = \Im (L^A)$ associated with $L^A$ must obey Dirichlet 
vanishing conditions, 
\bea
\label{8a2}
h^A  = 0 \quad \hbox{on} \quad \p \Sigma 
\eea
With the help of the following relation, 
\bea
\label{8a2b}
\bar \lambda \cdot \lambda -2 = { 4 \over |L^6|^2} \left (  (h^1)^2 
- \sum _{R=6}^{m+5} (h^R)^2 \right )
\eea
the second constraint of (\ref{1b2}) may be translated into the following inequality, 
which is strict in the interior of $\Sigma$,
\bea
\label{3a2}
(h^1)^2 > \sum _{R=6}^{m+5} (h^R)^2
\eea
The construction of \cite{Chiodaroli:2011nr} specializes to the case where 
$\Sigma$ has only one boundary component, i.e. $\Sigma$ is 
a the disk. The light-cone variables introduced here in (\ref{8a1}) and the 
conditions for regularity of (\ref{8a2}) and (\ref{3a2}) hold, 
however, for general Riemann surfaces $\Sigma$ with arbitrary numbers of 
boundary components.

\subsection{Regular solutions for the disk or upper-half-plane}
\label{resoldisk}

The simplest Riemann surface $\Sigma$ for which our half-BPS solutions can be constructed 
has one boundary and can be represented by a disk of unit radius.  In the following we 
conformally map the disk  to the upper half-plane where the boundary is given by the real line. 
The harmonic function $H$ is restricted by the regularity conditions to be of the form,
\bea
\label{1b3}
H(w,\bar w) =  \sum _{n=1}^N  \left ({ i \, c_n \over w - x_n} - { i \, c_n \over \bar w - x_n} \right )
\hskip 1in 
 c_n >0
\eea
The meromorphic light-cone functions $L^A$  may be 
parametrized by a finite number of simple (auxiliary) poles and take the form,
\bea
\label{8a4}
L^A(w) = \ell ^A _\infty + \sum^P _{p=1}  { \ell ^A _p \over w-y_p}
\eea
The positions of the poles $y_p$,  their residues $\ell ^A_p$, and the values 
$\ell ^A _\infty$ must all be real, and the index runs over $A=1,6,7,\ldots, m+5$.
Regularity of the solutions requires that all the zeros of $L^6$ and  $\p_w H$ be 
common, and that $L^6$ be regular at the poles of $\p_wH$. Therefore,  
$L^6$ may be expressed as follows,
\bea
\label{8a5}
L^6 (w) = i \tilde \ell^6_\infty {\prod_{n}^{N} (w-x_n)^2 \over \prod_{p}^P (w-y_p) } \p_w H  (w)
\eea
Matching between the expressions for $L^6$ given in (\ref{8a4}) and (\ref{8a5}) requires 
the number of auxiliary poles $P$ to be related to the number of physical poles $N$,  by the 
relation $P=2N-2$, and the constants at $\infty$ to obey the relation $\ell ^6 _\infty = \tilde \ell^6 _\infty
\sum _{n=1}^N c_n$. Finally, regularity of the $\l^A$ at the auxiliary poles, and constancy of the 
sign of $h_1<0$,  require the relation,
\be 
\label{8a6}
\ell^1_p =  \left ( \sum_{R=6} ^{m+5} \ell_p^R \ell_p^R  \right )^\half
\ee
As a result of the Schwarz inequality, condition (\ref{3a2}) is then automatically satisfied.
\begin{figure}[h]
\begin{center}
\includegraphics[scale=.43]{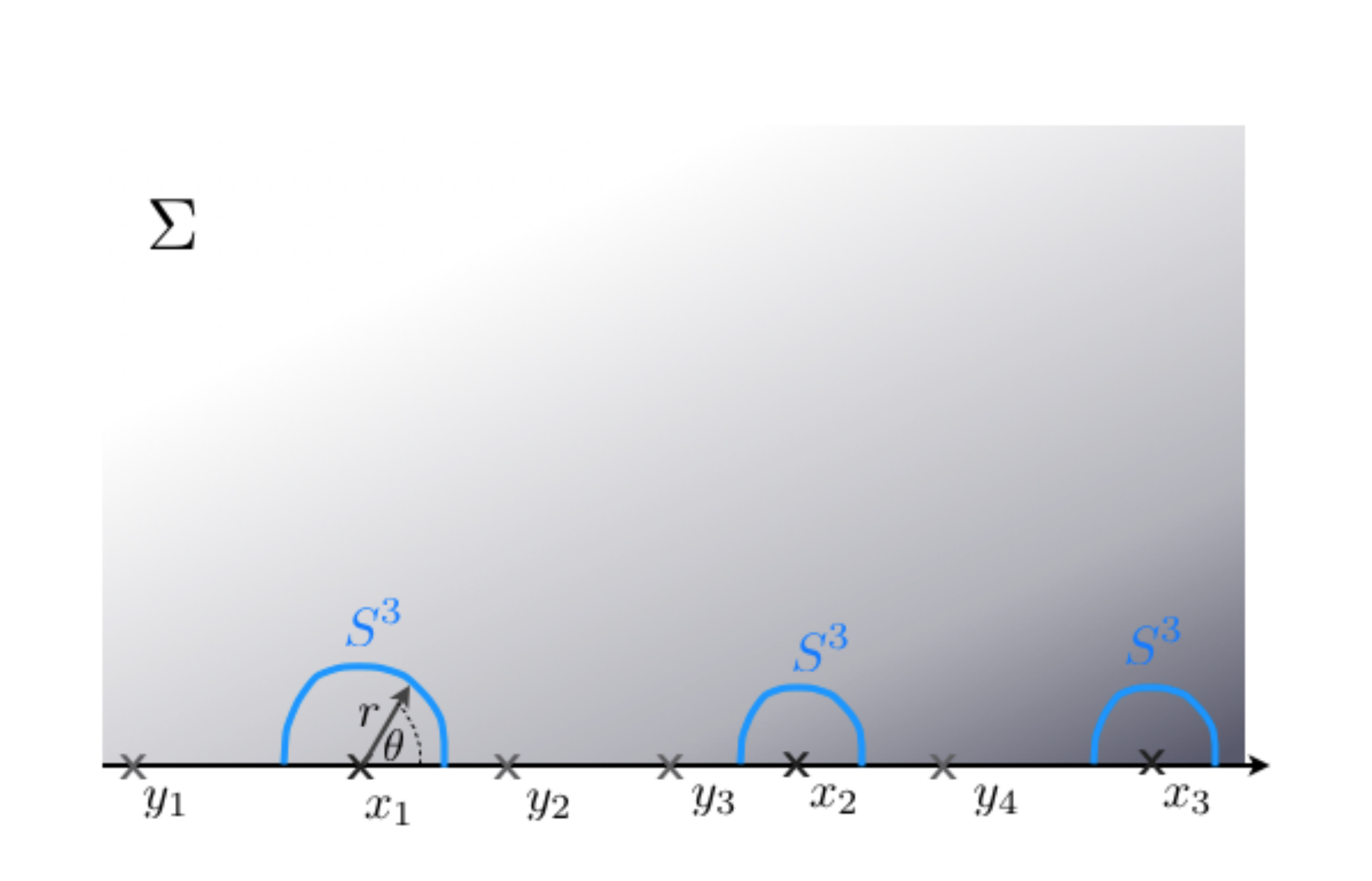} 
\caption{\small Holographic string-junction solution with $N=3$ asymptotic $AdS_3 \times S^3$ 
regions localized at  the poles $x_n$, with $n=1,2,3$ and $P=4$ auxiliary poles at $y_p$, with $p=1,2,3,4$.
Each homology $S^3$ is realized as an $S^2$ fibered over a line interval $C(x_n)$ represented here by
a half-circle in blue.}
\label{draft-fig4}
\end{center}
\end{figure}

Near the location $x_n$ of a pole  of $H$ the metric takes the following form,
\begin{equation}
\label{8a20}
 ds^2 \sim R_n ^2\left (  {dr^2 \over r^2} 
 +{r_n^2 \over r^2}  ds_{AdS_2}^2+ d\theta^2 + \sin^2\theta \, ds_{S^2}^2 
 \right )+ \cO (r^2)\nonumber
\end{equation}
where $w= x_n +  r e^{ i \theta}$, as depicted in Figure \ref{draft-fig4}. 
As $r \to 0$ the geometry approaches that of  
an asymptotic $AdS_3\times S^3$ region. The radius parameters $R_n$ and $r_n$
at a pole $x_n$ are determined by the asymptotic AST charge $Q_n^A$ as follows,\footnote{These
results correspond to formula (7.14) of \cite{Chiodaroli:2011nr}, with $\mu^A_n$ re-expressed
in terms of $Q_n^A$ using (7.15).}
\bea
\label{8a21}
R_n^4 = { Q_n \cdot Q_n \over 4 \pi^4}
\hskip 1in 
r_n^2 = { 16 \pi^4 c_n^2 \over Q_n \cdot Q_n}
\eea
The charges $Q^A_n$ may be expressed directly in terms of the 
parameters of the solution by, 
\begin{equation}
\label{asycha}
Q^A _n  
\equiv  \oint _{(S^3)_n} dB^A
=  \sqrt{2} \pi   \int _{C(x_n)} \!\! dw \,   \partial_w H \, \lambda ^A + {\rm c.c.} 
= 2 \sqrt{2} \pi^2 c_n \p \lambda ^A(x_n)
\end{equation}
The asymptotic charge is supported  by an $S^3$ which is realized here by fibering 
an $S^2$ over small semi-circle $C(x_n)$ surrounding the pole $x_n$, as depicted in Figure \ref{draft-fig4}.

\sm

As explained in detail in \cite{Chiodaroli:2009yw} the holographic interpretation of these 
solutions  as interface theories  proceeds as follows:  each region near a pole of $H$  
produces a holographic boundary component on which a CFT defined on a half-space lives. 
The boundaries of these half-spaces  are glued together along a one-dimensional interface 
which is given by the boundary  of the $AdS_2$ factor.

\sm

The counting of moduli of the space of solutions on the disk with $N$ poles is presented 
in Table 1, 
{\small
\begin{table}[htdp]
\begin{center}
\begin{tabular}{cc}
parameters  & number\\
$x_n$ & $N$ \\
$c_n$ & $ N $ \\
$y_p$  & $2N-2$ \\
$l_p^R$ & $(m-1)(2N-2)$ \\
$l^A_\infty$ & $m+1$ \\
$SL(2,R)$& $-3$
\end{tabular}
\caption{Moduli counting for the half-plane yielding a total of $2(m+1)N -m-2$.}\label{disccount}
\end{center}

\end{table}%
}
where the total number of moduli of the solution is found to be
$2(m+1)N-m-2$.
This counting agrees with the corresponding counting of the number of physical parameters.   
We have $(m+2)(N-1)$ charges, where the number of independent asymptotic charges 
(\ref{asycha}) is reduced  from  $N$ to $N-1$ because of charge conservation. 
In addition there are the $m$ scalars whose values  are not fixed by the attractor  
mechanism in each asymptotic region giving $mN$ parameters. The total number 
of physical parameters agrees with the number of moduli given  in Table \ref{disccount}.

\newpage

\section{Half-BPS solutions on the annulus}
\setcounter{equation}{0}
\label{sec3}

The aim of the present paper is to construct  string-junction solutions for Riemann 
surfaces $\Sigma$ of general topology and arbitrary moduli. In this section, we shall take a 
first step in this direction, and consider the simplest case of a surface $\Sigma$ with two 
disconnected boundary components, namely an 
annulus. We shall describe the construction of the annulus solutions in detail here since it
provides a very explicit example of the methods that will be used to construct the solutions 
for general Riemann surface $\Sigma$ in Section \ref{sec6}.

\sm

The annulus has a single imaginary modular parameter $\tau= i t$ where $t$ is real and 
may be taken to be positive. We introduce complex coordinates $w, \bar w$, subject to 
the identification $w \sim w+ 1$, so that $\Sigma$ may be represented by a rectangular 
domain in the complex plane,
\bea 
\label{3a0}
\Sigma = \left \{ w \in \mathbf{C}, ~~ 0 \leq \Re (w) < 1, ~~ 0 \leq \Im(w) \leq {t \over 2}  \right \} 
\eea 
The two disconnected boundary components of $\p \Sigma$ are 
specified respectively by $\Im (w)=0$ and $\Im (w) = t/2$.  As is customary, 
we make use of the compact double Riemann surface $\hat \Sigma$,  which is a torus, and is defined by,
\bea
\label{3a1}
\hat \Sigma = \left \{ w \in \mathbf{C}, ~~ 
0 \leq \Re (w) < 1, ~~ -{ t \over 2}  < \Im(w) \leq {t \over 2}  \right \} 
\eea
together with the identifications $w \sim w+1$ and $w \sim w +it$.
Under the anti-conformal involution of complex conjugation $\phi (w) = \bar w$,  we may 
identify $\Sigma$ as the coset $\Sigma = \hat \Sigma / \phi$.

\subsection{Solving for  $H$ and $h^A$}

The constructions of the real harmonic functions $H$ and $L^A= \Im (L^A)$, used in the light-cone parametrization 
of the local solution, will proceed in parallel to one another, and may be systematically derived in all generality. 

\sm

One key requirement is that  we have $H =h^A=0$ on $\p \Sigma$, following (\ref{8a2}).
A second key requirement is that $H$ must be positive inside $\Sigma$, so that 
it cannot have poles in the interior of  $\Sigma$.
Thus, the poles of $H$ must be distributed on the two boundary components of $\Sigma$,
and will be denoted respectively by $x_n$ and $\tau/2 +x' _\np$ where the parameters  $x_n$ and $x'_\np$
are real, and the corresponding indices run over $n=1 \ldots, N$ and $\np = 1 , \ldots, N'$. 

\sm

To specify the meromorphic functions $L^A$, it will be convenient to first construct their 
associated real harmonic functions $h^A = \Im (L^A)$.  In view of the strict inequality (\ref{3a2}) 
in the interior of $\Sigma$, we see that $h^1$ cannot vanish inside $\Sigma$, and every pole of 
$h^R$ must be a pole of $h^1$. As a result, all the poles of $h^1$, and thus all the poles of $h^A$, 
must be on the boundary of $\Sigma$. These {\sl auxiliary poles} will be denoted respectively by $y_p$ and 
$\tau/2 +y'_\pp$, where the parameters $y_p$ and $y'_\pp$ are real, and the indices run over
 $p=1,\ldots, P$, and $\pp=1, \ldots, P'$.
 
 \sm
 
In view of their single-valuedness on the double surface $\hat \Sigma$,
the harmonic functions $H$ and $h^A$ must be doubly periodic, namely with periods $1$ and $\tau=it$.
They are uniquely, but {\sl formally}, obtained by summing over the contributions of all respective poles, 
\bea
\label{3a4}
H& \sim & - 2 \, \Im \sum _{ (r,s) \in \bZ^2 } \left ( \sum _{n=1}^N { c_n \over w-x_n - r - s \tau}
+ \sum _{\np=1}^{N'} {  c'_\np \over w-x'_\np - r - s \tau} \right ) 
\no \\
h^A& \sim &  \Im \sum _{ (r,s) \in \bZ^2 } 
\left ( \sum _{p=1}^P { \ell ^A _p \over w - y_p - r - s \tau}
+ \sum _{\pp=1}^{P'} { \ell '^A _\pp \over w-y'_\pp - r - s \tau} \right ) 
\eea
The sums over integers $(r,s)$ are  logarithmically divergent and may be 
regularized by adding functions linear and constant in $w$. The corresponding 
regularized sums are readily recognized to be related to the Weierstrass $\zeta (w)$ function. 
For our purposes, it will be more convenient to work with the following function, 
\bea
\label{3a5}
\zeta _0 (w) \equiv { \p_w \tet _1 ( w  ) \over \tet _1 ( w )}
+{2 \pi i w \over \tau}
\eea
where $\tet _1 (w) = \tet _1(w|\tau)$ is the Jacobi $\tet$-function, satisfying the periodicity relations,
\bea
\label{3a5a}
\tet_1 (w + 1) & = & - \tet _1 (w)
\no \\
\tet_1 (w + \tau ) & = & - e^{-i\pi (2w+\tau)} \tet _1 (w)
\eea
The functions $\zeta (w)$ and $\zeta _0(w)$ differ from one another by a term linear in $w$, 
which has been chosen so that $\zeta _0(w)$ is periodic in $w$ with imaginary period $\tau$.  
The remaining non-trivial monodromy of $\zeta _0 (w)$  is then given by,  
\bea
\label{3a6}
\zeta _0 (w+1) = \zeta _0 (w) + { 2 \pi i \over \tau}
\eea
The harmonic function $h_0 (w)\equiv -  \Im \, \zeta_0(w)$ enjoys the following properties, 
\begin{enumerate}
\itemsep=0in
\item $h_0 =- \Im \, \zeta_0$ vanishes on $\p \Sigma$, namely when $w \in \bR$ 
 and when $w-\tau/2 \in \bR$;
\item $h_0 = - \Im \, \zeta _0  > 0$ in the interior of $\Sigma$.
\end{enumerate}

\sm

Using the above properties, we are now in a position to write precise formulas 
for the harmonic  functions $H$ and $h^A$ (instead of the formal expressions 
of (\ref{3a4})), and we find, 
\bea
\label{3a3}
H& = &  
- 2 \, \Im  \sum _{n=1}^N  c_n \, \zeta _0 (w-x_n) 
- 2 \, \Im \sum _{\np=1}^{N'}   c'_\np \, \zeta_0 ( w' + x'_\np) 
\no \\
h^A& = &
\Im \sum _{p=1}^P  \ell ^A _p \, \zeta _0 (w-y_p)
+ \Im \sum _{\pp=1}^{P'} \ell '^A _\pp \, \zeta _0 ( w'+ y'_\pp) 
\eea
Throughout, we shall use the shorthand notation, 
\bea
\label{3a7}
w'= {\tau \over 2} - w
\eea 
The vanishing of $H$ and $h^A$ on both boundary components of $\Sigma$
requires the residues $c_n,\ell^A_p,c'_\np , \ell '^A_\pp$ to be real. 
Positivity of $H$ throughout the interior of $\Sigma$ in turn requires, 
\bea
\label{3a8}
c_n > 0  & \hskip 1in & n=1, \dots , N
\no \\
c'_\np > 0 &&  \np=1, \dots ,N' 
\eea

\subsection{Parametrizing $L^A$}

The meromorphic functions $L^A$, for $A=1,6,7, \ldots, m+5$, 
may be inferred from the form of the harmonic functions $h^A$, and we find, 
\bea
\label{3c1}
L^A = L_0^A + \sum _{p=1}^P  \ell ^A _p \, \zeta _0 (w-y_p)
+ \sum _{\pp=1}^{P'} \ell '^A _\pp \, \zeta _0 ( w'+ y'_\pp) 
\eea
where $L_0^A$ are real constants whose values are undetermined by $h^A$.
Whereas the harmonic functions $h^A$ have been constructed to be single-valued on the annulus, 
the corresponding meromorphic functions $L^A$ may have additive monodromy, 
\bea
\label{3c2}
L^A (w+1) = L^A (w) + K ^A 
\hskip 1in K ^A = { 2 \pi i \over \tau} 
\left ( \sum _{p=1}^P \ell ^A _p - \sum _{\pp=1}^{P'} \ell '^A _\pp \right )
\eea
The light-cone construction requires $L^6$ to be single-valued on the boundary $\p \Sigma$,
so that we must impose the following relation,
\bea
\label{3c2a}
0 = K ^6 = \sum _{p=1}^P \ell ^6 _p - \sum _{\pp=1}^{P'} \ell '^6 _\pp
\eea
As a result, $L^6$ is a doubly periodic function with periods 1 and $\tau$.
The absence of second order auxiliary poles in $\l^2$, $\l^6$ at both boundary 
components requires,
\bea
\label{3c3}
\ell ^1 _ p = \left ( \sum _{R=6}^{m+5} (\ell _p ^R)^2 \right  )^\half
\hskip 1in 
\ell '^1 _ \pp = \left ( \sum _{R=6}^{m+5} (\ell _\pp '^R)^2 \right  )^\half
\eea
both square roots having a positive sign.
The positions $y_p$ and $\tau/2+ y'_\pp$ of the auxiliary poles and the residues 
$\ell ^A_p$ and $\ell '^A_\pp$ of the functions $L^A$ for $A=7,8, \ldots, m+5$ are 
free parameters of the solutions, and may be specified at will. The residues  
$\ell ^1_p$ and $\ell '^1_\pp$ are then deduced using (\ref{3c3}), as well as the 
values of $\ell^6_p$ and $\ell '^6_\pp$. Thus, it remains to determine 
$\ell^6_p$ and $\ell '^6_\pp$. 

\subsection{Determining  $L^6$}
\label{threethree}

The meromorphic function $L^6$ is single-valued on the torus $\hat \Sigma$.
Also, as a result of the regularity conditions in the light-cone parametrization, 
the function $L^6$ is required to have precisely the same zeros as $\p_w H$.
Since the numbers of poles of $L^6$ is $P+P'$, and the number of (double) 
poles of $\p_wH$ is $N+N'$, it follows that we must have, 
\bea
\label{3b1}
P+P'=2N+2N'
\eea
The meromorphic function $\cN$ on $\hat \Sigma$ defined by,
\bea
\label{3b2}
L^6 = { \p_w H \over \cN}
\eea
has prescribed zeros at $y_p$ and $\tau/2+y'_\pp$, and prescribed (double) 
poles at $x_n$ and $\tau/2 + x'_\np$. As such, $\cN$ is unique
up to a multiplicative complex  constant $\cN_0$. 
As a result, we have an alternative product formula for $\cN$ directly 
in terms of theta functions, 
\bea
\label{3b3}
\cN(w) = \cN_0 e ^{i \pi \mu w} 
{\prod_{p=1}^P \tet_1(w-y_p  ) \prod_{\pp=1}^{P'} \tet_1(w' + y'_\pp )
\over \prod_{n=1}^N \tet_1(w-x_n  )^2 \prod_{\np=1}^{N'} \tet_1(w' + x'_\np )^2 }
\eea
where $\mu$ is a constant, which remains to be determined. The factor 
$|\cN_0|$ enters  $L^6$ multiplicatively and, in light of the discussion of 
\ref{twothree}, produces an $SO(5,m)$ boost on $\l^\pm$. As a result, 
the presence of $|\cN_0|$ may be undone by an overall $SO(5,m)$ transformation.
Henceforth, we shall set $|\cN_0|=1$ without loss of generality. The phase 
of $\cN_0$ needs to be retained to render the function $L^6$ real on the boundary $\p \Sigma$,
as will be carried out below.

\sm

Next, we implement the  requirement $\Im L^6(w)=0$ for $w \in \p \Sigma$.
Since $\p_w H(w)$ is purely imaginary for $w \in \p \Sigma$, and $L^6(w)$ is real there, 
$\cN(w)$ must be purely imaginary for $w \in \p \Sigma$. For real $x$, 
the function $\tet_1 (x | \tau)$ is real and we have,\footnote{This relation is proven using
$\tet _1 (x + \tau/2) = \tet _4 (x)\, \exp ( i \pi /2 - i \pi x + \pi t/4) $
and  $\tet _4 (x) > 0$ for real $x$.}
\bea
\label{3b4}
\tet _1 ( \tau/2 -x) = \Big | \tet _1 ( x + \tau/2 ) \Big | \, e^{i \pi /2 + i \pi x}
\eea
Using these relations, we deduce the phase of $\cN(w)$ for all points on $\p \Sigma$, 
\bea
\label{3b5}
\cN (x) & = & |\cN (x) | \, \exp i \pi \left \{ x(\mu - 2N' +P') + {P'\over 2} 
+ 2 \sum _\np x'_\np - \sum _\pp y' _\pp \right \}\no
\\
\cN (x + \tau/2) & = & | \cN (x + \tau/2) | \exp i \pi \left \{ x(\mu + 2N -P) + {P\over 2} 
- 2 \sum _n x_n + \sum _p y _p \right \}
\eea
Requiring constancy of the phase on each boundary determines $\mu$ by, 
\bea
\label{3b6}
\mu = 2N'-P'=P-2N
\eea
The second equality coincides with (\ref{3b1}). The remaining 
phases are now constant throughout each boundary component, and may be 
absorbed by the phase of $\cN_0$ provided the phases on both boundary components
coincide modulo integers, giving the final condition, 
\bea
\label{3b7}
2 \sum _n x_n +  2 \sum _\np x'_\np -  \sum _p y _p - \sum _\pp y' _\pp 
\equiv 0 \qquad  ({\rm mod} ~ 1)
\eea
Here we have used the fact that $(P'-P)/2$ is an integer since
$P+P'$ is even by (\ref{3b1}). 

\sm

\begin{figure}[h]
\begin{center}
\includegraphics[scale=.35]{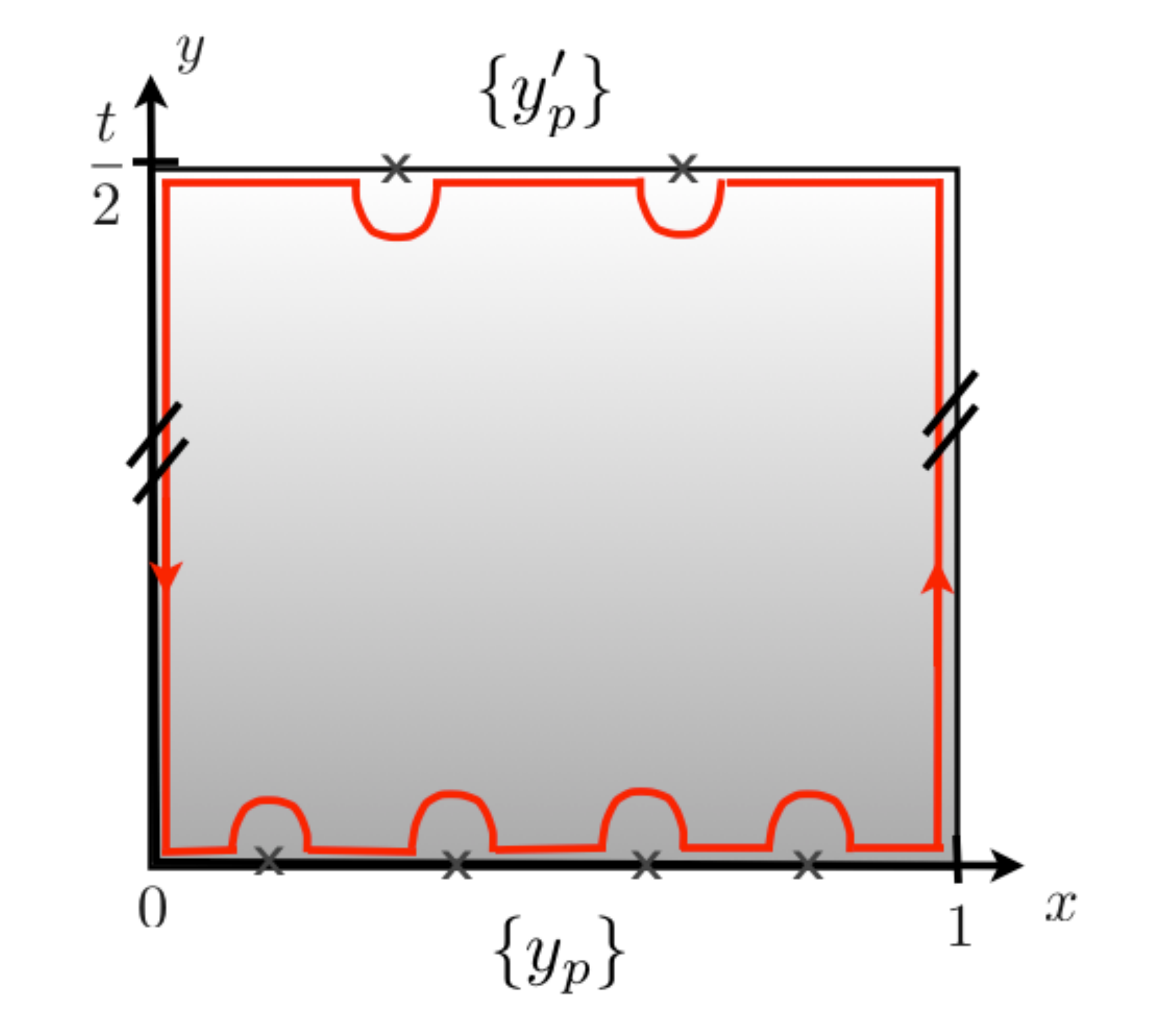} 
\caption{Contour (in red) around which to integrate $L^6$ to establish residue relation.}
\label{annfig4}
\end{center}
\end{figure}

The monodromies of $\cN$ may be computed using the relations of (\ref{3a5a}),
and we find, 
\bea
\label{periodn}
\cN (w+1) & = & e^{i \pi \mu } \cN (w)
\no \\
\cN ( w + \tau) & = &  \cN(w) 
\eea
Single-valuedness of $\cN$ on $\hat \Sigma$ requires  $\mu$, and thus $P$ and $P'$, to be even.
Having determined $L^6$, we are now in a position to extract its 
residues at the poles $y_p$ and $y'_\pp$, and we find, 
\bea
\label{3b8}
\ell ^6 _p = { \p_w H(y_p) \over \p_w \cN (y_p)}
\hskip 1in 
{\ell '}^6 _\pp = - { \p_w H(y'_\pp) \over \p_w \cN(y'_\pp)}
\eea
These residues do not manifestly exhibit the absence of monodromy relation $K^6=0$ of (\ref{3c2a}).
It may be proven using single-valuedness of $\cN$ and thus $L^6$,  term by term in the sums over $n, n'$ in $H$, 
and the vanishing integral of $L^6$ over the contour in red of Figure \ref{annfig4}.

\subsection{Proof of the regularity}

To obtain globally regular solutions, it remains to enforce the strict inequality of (\ref{3a2})
in the interior of $\Sigma$. We shall now prove that (\ref{3a2}) holds using the expressions 
for $h^A$ given in (\ref{3a3}), the expressions for the residues $\ell ^1 _p$ and ${\ell '}^1 _\pp  $ 
given in (\ref{3c3}), as well  as the strict positivity of the functions $- \Im \, \zeta _0 (w-y_p)$
and $- \Im \, \zeta _0 (w'+y'_\pp)$ for $w$ in the interior of $ \Sigma$.
We shall  use the following shorthand notation, 
\bea
\label{3d1}
W_p & = &  - \Im \, \zeta _0 (w-y_p) 
\no \\
W _\pp & = & - \Im \, \zeta _0 (w'+ y' _\pp)
\eea
In the interior of $\Sigma$, the strict inequalities $W_p>0$ and $W_\pp >0$ hold.
Using the above notation, inequality (\ref{3a2}) will be satisfied 
provided the following quantity,
\bea
\label{3d2}
h^A h_A = \left ( \sum _p \ell ^1 _p W_p + \sum _\pp \ell '^1 _\pp W_\pp \right )^2 -
\sum _{R=6}^{m+5} \left ( \sum _p \ell ^R _p W_p + \sum _\pp \ell '^R _\pp W_\pp \right )^2 
\eea
is strictly positive. 
The structure of the equation may be clarified by introducing vector notation, 
and by explicitly solving for $\ell^1_p$ and ${\ell '} ^1 _\pp$ using (\ref{3c3}),
\bea
\label{3d3}
\ell ^1 _p = | \vec{\ell} _p | \, & \hskip 1in & 
\vec{\ell}_p = ( \ell _p ^6, \ell^7 _p, \ldots, \ell ^{m+5}_p)
\no \\
\ell '^1 _\pp = | \vec{\ell} '_\pp | & \hskip 1in & 
\vec{\ell }' _\pp = ( {\ell '} _\pp ^6, {\ell '} ^7 _\pp, \ldots, {\ell ' }^{m+5}_\pp)
\eea
and by separating terms according to their boundary dependence, 
\bea
\label{3d4}
h^A h_A & = & 
~~ \sum _{p,q} 
\Big ( | \vec{\ell}  _p| \, | \vec{\ell} _q | -\vec{\ell} _p \cdot \vec{\ell} _q \Big ) W_p W_q
+ \sum _{\pp, \qp} 
\Big ( | \vec{\ell}'  _\pp| \, | \vec{\ell} ' _\qp | -\vec{\ell}' _\pp \cdot \vec{\ell}' _\qp \Big ) W_\pp W_\qp
\no \\ &&
+ 2 \sum _{p,\pp} 
\Big ( | \vec{\ell}  _p| \, | \vec{\ell}' _\pp | -\vec{\ell} _p \cdot \vec{\ell}' _\pp \Big ) 
W_p W_\pp
\eea
By the Schwarz inequality, the sum is positive or zero term by term,
at all points $ w \in \Sigma$. 

\sm

The Schwarz inequality implies an even stronger result: if $h^Ah_A$ vanishes at
any interior point of $\Sigma$, then it must vanish identically throughout $\Sigma$.
To establish this result, we assume that  $h^A h_A$ vanishes at a single interior 
point of $\Sigma$. This would imply the
vanishing of every term in the three sums. Since we have $W_p>0$ and 
$W_\pp>0$ strictly  in the interior of $\Sigma$, this in turn would require that all 
vectors $\vec{\ell}_p$ and $\vec{\ell}' _\pp$ must be collinear to one single vector.
But when this is the case, the function $h^A h_A$ vanishes throughout $\Sigma$.
Note that, for fixed poles $x_n, x'_\np, y_p, y'_\pp$ and residues $c_n, c'_\np$, the collinearity 
condition gives us a simple linear criterion for the degeneration of the solutions.

\subsection{Monodromy structure of $\l^A$}
\label{monola}

The monodromy of $L^A$, for $A=1,6,7, \ldots, m+5$, under the shift $w \to w+1$
was derived in (\ref{3c2}) and is encoded in the quantities $K ^A$. 
Single-valuedness of $L^6$ implies $K ^6=0$. The function $H$ is, of course,
single-valued. Non-trivial monodromies $K ^A$ imply non-trivial transformation  
laws for the meromorphic functions $\lambda ^A$, which we shall now obtain. 

\sm

The monodromy relations on $L^A$ of (\ref{3c2})  imply monodromy relations for $\lambda ^A$ 
namely,\footnote{We shall extend the dot notation to inner products of the type 
$K \cdot \l = K ^1 \l^1 - K ^R \l^S \delta _{RS}$  where the 
summation indices runs over the values $R,S=7,8,\ldots, m+5$.}
\bea
\label{3e2}
\l^A (w+1) & = & \l^A (w) + 2 K^A \l^- 
\no \\
\l^+ (w+1) & = & \l^+ (w) - 2 K \cdot \l - 2 (K \cdot K) \l^-
\no \\
\l^- (w+1) & = & \l^- (w)
\eea
where the index $A$ on the first line runs over the values $A=1,7,8, \ldots, m+5$.
A matrix form manifestly exhibits the multiplicative structure of the monodromy on $\lambda ^A$, 
\bea
\label{3e3}
\l (w+1) = M \l (w) 
\eea
with $\lambda$ and $M$ given by,
\bea
\label{3e4}
\lambda = \left ( \matrix{ \l^+ \cr \l^1 \cr \l^R  \cr \l^- } \right )
\hskip 1in 
M = \left ( \matrix{ 
1 	& - 2 K ^1 	& + 2 K ^R 	&  - 2 K \cdot K 	\cr
0 	& 1			&  0				& 2 K ^1 		\cr
0	& 0 			& I 				& 2 K ^R		\cr
0	& 0			& 0 		  		& 1 \cr} \right )
\eea	    
The matrix $M$ is naturally the exponential of a simpler matrix, $M = \exp \cM$, where 
\bea
\label{3e5}
\cM = \left ( \matrix{ 
0 	& - 2 K^1 	& 2 K^R 	& 0 	\cr
0 	& 0			& 0     	&  2 K^1 		\cr
0	& 0 			& 0 		& 2 K ^R		\cr
0	& 0			& 0 		& 0 \cr} \right )
\hskip 1in 
\eta = \left ( \matrix{ 
0 	& 0	& 0 	& 1 	\cr
0 	& 1			& 0     	&  0 		\cr
0	& 0 			& -I 		& 0		\cr
1	& 0			& 0 		& 0 \cr} \right )
\eea
In light-cone coordinates, the $SO(2,m)$ metric  $\eta$ takes the form given in (\ref{3e5}).
One may readily check the relations $\cM^t \eta + \eta \cM=0$ and $M^t \eta M = \eta$,
from which we conclude that the matrix $M$ belongs to the real group $SO(2,m)$. As a result, the 
space-time metric of the solution is invariant, while the scalar fields $V$ and the 2-form $B$
transform non-trivially under the monodromy $M$. Two different matrices $\cM, \cM'$ 
of the form (\ref{3e5}) commute with one another for any assignments of their entries 
$K$ and $K'$. This is as expected since $M$ must form a representation of the 
Abelian group of homology cycles.

\sm

Since the scalar field $V$ and the 2-form field $B$ transform non-trivially under monodromy
on $\Sigma$, these space-time  fields are multiple-valued and thus not strictly
speaking supergravity solutions. Quantum mechanically, however, anomalies lead
to reducing the continuous symmetry group $SO(5,m)$ to its discrete U-duality subgroup
$SO(5,m;\bZ)$. As a result, the scalars should really live on the coset 
$SO(5,m;\bZ) \backslash SO(5,m;\bR) /SO(5) \times SO(m)$, and a quantum solution
in string theory allows for monodromy of the fields that lie in $SO(5,m;\bZ)$.

\sm

In order to realize these allowed monodromy in the U-duality group $SO(5,m;\bZ)$,
we must require that the monodromy matrix $M$ of (\ref{3e2}) belong to 
$SO(5,m;\bZ)$, {\sl up to a conjugation by a constant element $g$ of} $SO(5,m;\bR)$,
\bea
g M g^{-1} \in SO(5,m;\bZ) \hskip 1in g \in SO(5,m;\bR)
\eea
The conjugation $g$ is in general required on a coset to account for the precise
embedding of $SO(5,m;\bZ) $ in $SO(5,m;\bR)$, and the realization thereof on the fields.
To see what this condition implies, we choose the basis in which $g=I$, and we see
that it suffices to take $K^A \in \bZ$. 
There exists,  however, also a more delicate class of solutions for $M$, in
which $2K^A \in \bZ$, and $2K \cdot K \in  \bZ$, so that 
the lattice of $K^A/2$ must be integral and even.

\subsection{Counting moduli and physical parameters}\label{countanu}

The counting of moduli and physical parameters of the solution for the annulus 
generalizes the  case of the disk given  in Section \ref{resoldisk}. The moduli of the solutions are 
counted in Table 2,

\begin{table}[htdp]
\begin{center}
\begin{tabular}{cc}
parameters  & number of moduli\\
$x_n, x'_\np$ & $N_{tot}=N+N'$\\
$c_n,c'_\np$ &$ N_{tot}$\\
$y_p, y'_p$  & $2N_{tot} = P+P'$\\
$l_p^R, l'_p{}^R$ &$ (m-1)2N_{tot}$\\
$l^A_\infty$ & $m+1$ \\
$t$& $+1$ \\
$U(1)$& $-1$ \\
$2\sum x=\sum y$& $-1$\\
\end{tabular}
\label{ancount}
\caption{Moduli counting for the annulus yielding a total of $2(m+1)N_{tot}+m$.}
\end{center}
\end{table}%

The last entry in Table 2 stands for the constraint (\ref{3b7}). 
Physical parameters are counted as follows. There are $(m+2)N_{tot}$ charges.  
The number of scalars whose values are not fixed by the 
attractor mechanism  is $mN_{tot}$. There are also $m$ axionic scalars, whose values are 
subject to shifts under the monodromy matrix $M$. We can associate these shifts with the 
presence of a D3-brane. Altogether one has $2(m+1)N_{tot}+m$
physical parameters in agreement with counting of the total number of  moduli presented in Table 2.

\newpage

\section{Simple regular annulus solution dual to BCFT}
\setcounter{equation}{0}
\label{sec5}

The simplest annulus solution has a single asymptotic $AdS_3 \times S^3$ region, 
and corresponds to a single pole for $H$ which we may choose to be on the lower boundary,
so that $N=1$ and $N'=0$.  Condition (\ref{3b1}) together with the fact that the 
number of auxiliary poles on either boundary has to be even leaves either $P=2, P'=0$ 
or $P=0, P'=2$. We shall analyze in detail the case $P=2$.  The case $P'=2$
produces qualitatively the same results.

\begin{figure}[h]
\begin{center}
\includegraphics[scale=.45]{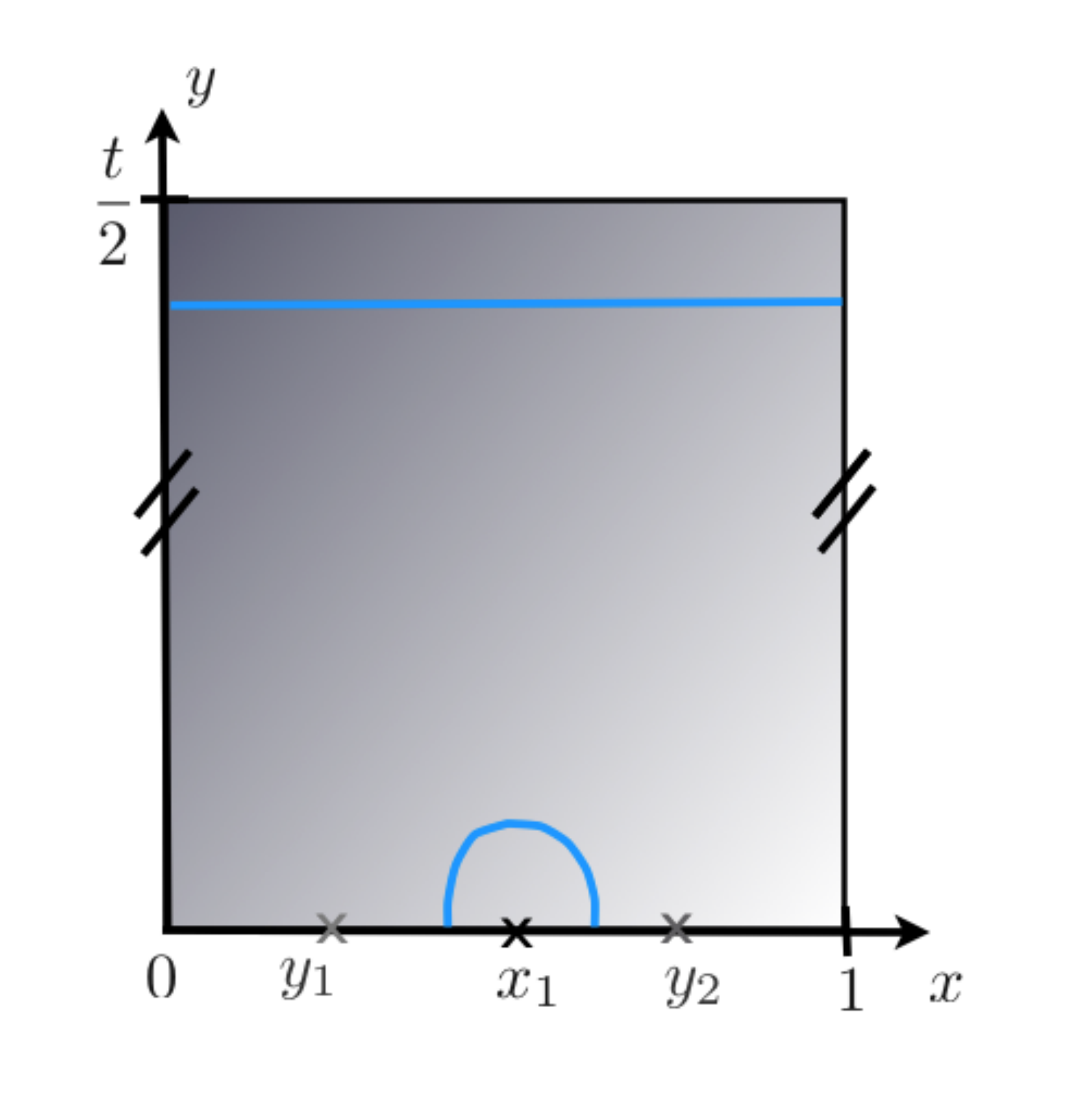} 
\caption{BCFT example of the annulus with $N=1,P=2$.}
\label{draft-fig5}
\end{center}
\end{figure}

The holographic boundary is given by a single half-space and hence this is an example 
of a non-singular holographic realization of a BCFT. Denoting the pole in $H$ by $x_1$
and the auxiliary poles by $y_1,y_2$, condition (\ref{3b7}) requires $2x_2=y_1+y_2$ 
up to the addition of an integer. The relevant functions then become, 
\bea
\label{anntheta}
H& = &  
- 2 c_1 \, \Im  \, \zeta _0 (w-x_1) 
\nonumber\\
L^A &=& L_0^A + \ell ^A _1 \, \zeta _0 (w-y_1)+ \ell ^A _2 \, \zeta _0 (w-y_2)
\nonumber \\
\cN(w) &=& \cN_0 
{  \tet_1(w-y_1  )  \tet_1(w-y_2  ) \over  \tet_1(w-x_1  )^2  } 
\eea
In light of the discussion given in Section \ref{threethree}, we set $ |\cN_0|=1$
without loss of generality.
Translation invariance on the torus allows us to set $x_1=0$, and $y_1=-y_2=y$. 
The asymptotic charges are defined as in (\ref{asycha}) by the contour integral along a 
semi-circle enclosing the pole $x_1$ on the boundary (see Figure \ref{draft-fig5}).   
For a simply connected $\Sigma$ deforming the contour integral would  imply that 
asymptotic charge for a solution with a single pole vanishes. For the annulus there is 
a non-contractible cycle. Consequently the asymptotic charge is not vanishing as can 
be checked by explicit calculation done in the next section.

\vskip -0.1in

\begin{figure}[h]
\begin{center}
\includegraphics[scale=.45]{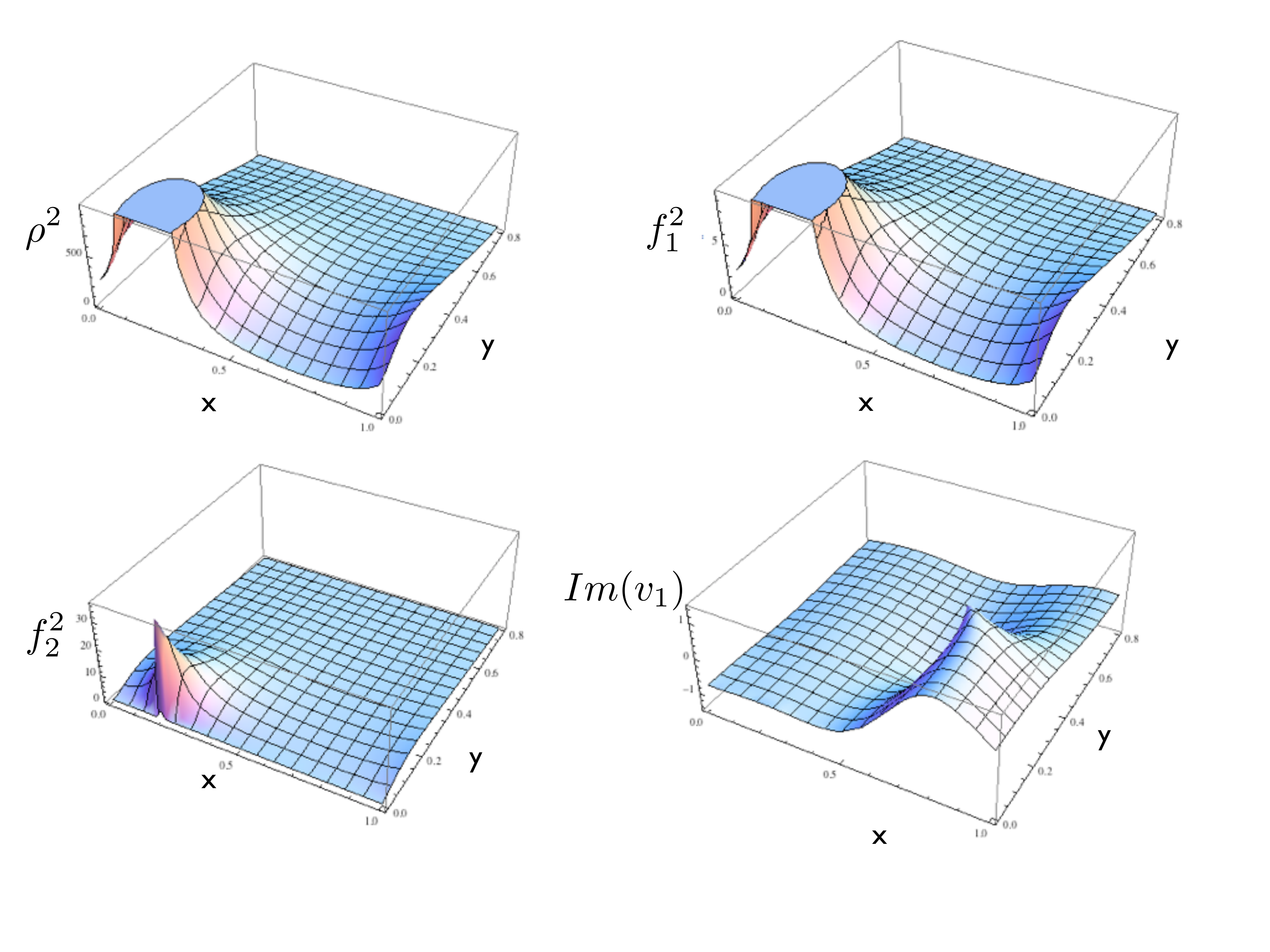} 
\caption{Plots for a $N=1, P=2$ example  with $t=1.6$ and $x_1=.2, y_1 =.25, y_2=.15$ 
and generic values for the other parameters} 
\label{annfig1}
\end{center}
\end{figure}

\vskip -0.1in 

From (\ref{anntheta}) explicit expressions for the metric factors, antisymmetric tensor 
fields and scalars can be obtained using the formulae presented in Section \ref{sec2b}. 
Here we present a plots for some generic values of the parameters. The metric factors 
are everywhere regular and the behavior of a component of the coset frame $\Im(v^1)$ 
indicates the existence of a non-trivial monodromy for the scalars as discussion in Section \ref{monola}.

\subsection{General formula for the Boundary Entropy}
\label{boundcalc}

Following \cite{Affleck:1991tk}, the boundary entropy 
$S_{{\scriptstyle \rm bcft}}$ is related to the ground state degeneracy 
$g_{{\scriptstyle \rm bcft}}$ of the degrees of freedom localized on the boundary by 
$S_{{\scriptstyle \rm bcft}}=  \ln g_{{\scriptstyle \rm bcft}}$. 
For two-dimensional CFTs the boundary entropy can be extracted from the entanglement 
entropy $S_A$ of a spatial domain $A$ which is an interval of length $L$ ending at the 
boundary \cite{Calabrese:2004eu},
\be
\label{saform}
S_A=  {c\over 6} \ln {L \over \ep} + S_{{\scriptstyle \rm bcft}}
\ee
Here $c$ is the central charge of the CFT and $\ep$ is a UV cutoff. 
In \cite{Ryu:2006bv,Ryu:2006ef} a holographic prescription to calculate the 
entanglement entropy was developed. The entanglement entropy can be calculated as follows,
\be
S_A={ {\rm Area}(\gamma_A)\over 4 G_N^{(d+1)}}
\ee
where $\gamma_A$ is the minimal surface in the bulk which on the boundary encloses 
the spatial domain $A$ and $G_N=G_N^{(d+1)}$ is the Newton constant in $d+1$ dimensions. 
In \cite{Chiodaroli:2010ur}, a generalization of this prescription was put forward
to the case of Janus solutions which are $AdS_2\times S^2$ fibrations over a Riemann 
surface $\Sigma$. The minimal surface is obtained by integrating over $S^2$ 
as well as $\Sigma$ giving in our case,\footnote{With the metric $ds^2 _\Sigma = 4 \rho^2 |dw|^2$, 
the volume form on $\Sigma$  is given by $4 \rho^2 d^2w$ where $d^2 w = d \Re(w) \, d \Im (w)$.}
\bea
\label{7c0}
S_A  &=& { 1 \over 4 G_N} \int _{\Sigma _\ep} d^2w \,  4 \rho^2 \,  f_2^2  \int _{S^2}\,  \omega_{S^2}
\no \\
&=& {\pi \over  G_N} \int _{\Sigma _\ep} d^2w \, |\p_w H |^2  \left ( \l \cdot \bar \l -2 \right )
\eea
where (\ref{7c1}) was used  to arrive at the second line. Here $\Sigma_\epsilon$  is a regularized
surface obtained from $\Sigma$ by cutting out a small half-disk around the pole of $H$ located at $w=x_1$. 
A very useful formula is obtained by converting $\lambda^A$ into the harmonic functions $h^A$
using (\ref{8a2b}), and $\p_wH$ into $\cN$ using (\ref{3b2}), and we have,
\bea
\label{7c9}
S_A = { 4 \pi \over G_N} \int _{\Sigma _\ep} d^2 w \, |\cN |^2 h^A h_A
\eea
where the summation in $A$ runs over $A=1,7,8, \ldots, m+5$.

\subsection{Calculation of the Boundary Entropy}

Applying the general formula (\ref{7c9}) for the entanglement entropy to the case of the BCFT 
described in (\ref{anntheta}), we find the following simplified expression, 
\bea
\label{sboundabd}
S_A = { 8 \pi \over G_N} \Big (
|\vec{\ell}_1 | \, | \vec{\ell}_2 | - \vec{\ell}_1 \cdot \vec{\ell}_2\Big ) \int _{\Sigma _\ep} 
d^2w \, |\cN|^2 h_0(w-y_1) h_0(w-y_2)
\eea
The dependence on the cutoff $\ep$ is extracted in Appendix \ref{chargedep}, and we find, 
\bea
\label{5a1}
S_A= {Q \cdot Q \over 8 \pi^2 G_N}  \ln  { 1 \over 2 \pi \ep} + S_{{\scriptstyle \rm bcft}}
\eea
where $Q^A$ are the charges of the asymptotic $AdS_3 \times S^3$ region, and the 
central charge $c$ of the bulk CFT is given by, 
\bea\label{cvaluean}
{c\over 6} ={ Q \cdot Q \over 8 \pi^2 G_N} 
\eea
This central charge is in accord with the Brown-Henneaux formula derived from the 
radius of the asymptotic $AdS_3$, 
\bea
\label{5a2}
{c \over 6} = {  \ell _3 \over 4 G_3} \hskip 1in {1 \over G_3} = {{\rm Vol} (S^3) \over G_N}
\eea
where $\ell_3$ is the $AdS_3$ radius, Vol$(S^3)$ the volume of the sphere, and $G_3$
is the effective 3-dimensional Newton constant. By inspection of (\ref{8a20}) and (\ref{8a21}),
we see that $\ell _3 = R_n$, and Vol$(S^3) = 2 \pi^2 R_n^3$, so that we recover 

\sm

We have succeeded in analytically evaluating the integral of (\ref{sboundabd}) 
for large enough modular parameter $t$. To do so, we make use of the {\sl uniform 
approximation} of $\tet_1(w)$ and $\zeta _0(w)$ in the domain given in (\ref{3a0}) 
for $\Sigma$ by
the function, 
\bea
\label{5a3}
\tet _1 (w ) & = & 2 e^{-\pi t /4}  \sin \pi w + \cO(e^{-2 \pi t}) 
\no \\
\zeta _0 (w) & = & \pi \cot \pi w + { 2 \pi w \over t} + \cO(e^{-2 \pi t}) 
\eea
In a lengthy calculation,
which has been deferred to Appendix \ref{asyment}, the finite part $S_{{\rm \scriptstyle bcft}}$ 
is computed analytically in $y$ and $t$ up to terms which are suppressed by powers of $e^{-\pi t}$. 
The final result  is given by,
\bea
\label{5a4}
S_{{\rm \scriptstyle bcft}} 
 = {4 \pi^2 \over G_N} \left ( |\vec{\ell}_1| \, |\vec{\ell}_2| - \vec{\ell}_1 \cdot \vec{\ell}_2 \right ) 
 \left (   {\pi t \over 3} - \half  - { 2 \over \pi t} \sin ^2 \pi y + { 4 \over \pi^2 t^2 } \sin ^4 \pi y \right )
 + \cO(e^{- \pi t})
\eea
The dependence on powers in $t$, and the $y$-dependence thereof, is exact.
There is some ambiguity, of course, in the splitting between the divergent part
and the boundary entropy, as a change in regulator $\ep$ will induce a 
change in $S_{{\rm \scriptstyle bcft}}$. But this change has characteristic $t$
and $y$-dependence, and may be clearly isolated.

\subsection{Dependence on the scalar monodromy}

For the simple BCFT discussed in the present section the additive  monodromy $K^A$  given in (\ref{3c2}) becomes
\be\label{admod}
K^A={2\pi\over t} \Big( l_1^A+ l_2^A\Big)
\ee
From which we can calculate the norm
\be\label{kasqu}
K_A K^A = {8\pi^2\over t^2} \left ( |\vec{\ell}_1| \, |\vec{\ell}_2| - \vec{\ell}_1 \cdot \vec{\ell}_2 \right ) 
\ee
Where we used the fact that $\vec{\ell}_1{}^{A} \vec{\ell}_{1} {}_{A} =\vec{\ell}_2{}^{A} \vec{\ell}_{2} {}_{A}=0$. 
Using (\ref{cvaluean}) and (\ref{a11}) the central charge of the BCFT can be expressed as follows
\be\label{cform}
c= {24 \pi^2 \over G_N}  \Big( |\vec{\ell}_1 | \, | \vec{\ell}_2 | - \vec{\ell}_1 \cdot \vec{\ell}_2\Big )    
|\hat \cN _0 |^2    \hat{h}_0(y)^2
\ee
It follows from  (\ref{kasqu}) and (\ref{cform}) that  for a annulus with finite modulus $t$ setting the monodromy vector $K^A$ to zero implies that the central charge  $c$ vanishes. This  supports  the interpretation of the BCFT as a near-horizon limit of a self-dual string ending on a three-brane.  A vanishing monodromy implies the absence of a three-brane   and consequently the self-dual string has nowhere to end. In our  supergravity solution this is reflected by the fact that the a solution dual to a BCFT with $N=1$ and vanishing $K^A$ cannot be regular. On the other hand, solutions with $N>1$ corresponding to interfaces or junctions of self-dual strings exist with vanishing monodromy $K^A$.

\subsection{Degeneration of the annulus solution}

Next, we derive the asymptotic behavior of the annulus solutions of Sections \ref{sec3} and \ref{sec5} 
for large $t$, namely when the inner boundary component is shrunk. The nature of this degeneration
will help in understanding the connection with other solutions, such as the $AdS_2$-funnel
introduced in \cite{Chiodaroli:2011nr,Chiodaroli:2011fn}.
In taking this degeneration limit, we shall keep fixed the physical data in the asymptotic 
regions, namely the charges and values of the un-attracted scalars. We should expect to find the disk
solution of Section \ref{resoldisk} to leading order, as well as physically interesting sub-leading corrections
to the disk solution.

\sm

The approximations to be used here are the ones of (\ref{5a3}) that allowed us to derive the boundary 
entropy in a uniform expansion valid exactly in $t,y$, up to exponential corrections of the form $e^{-\pi t}$.
To connect the limiting solutions with those of the disk, it will be convenient to map the degenerated 
annulus onto the upper half-plane (in terms of which the disk solution was previously formulated)
by introducing the following change of coordinates,
\bea 
\label{5b1}
z = - \cot \pi w \, \qquad & \quad u_n = - \cot \pi x_n \quad & \qquad v_p = - \cot \pi y_p
\no \\
z' = - \cot \pi w' \qquad & \quad u'_{n'} = - \cot \pi x'_{n'} \quad & \qquad v'_{p'} = - \cot \pi y'_{p'}
\eea
Note that $u_n, u'_{n'}, v_p$, and $v'_{p'}$ are real.
The different regions of the annulus may  be described by the following 
coordinate regions, 
\bea
\label{5b2}
\Sigma _{\rm lower} & = & \Big \{ 0 \leq \Im (w) \ll t/2 \Big \}  
\no \\
\Sigma _{\rm center} & = & \Big \{ | \Im (w) - t/2) | \ll t/2 \Big \} 
\no \\
\Sigma _{\rm upper} & = & \Big \{ 0 \leq  \Im (w') \ll t/2 \Big \}
\eea
Suitable coordinates in each region are respectively $z$, $w$, and $z'$. 

\begin{figure}[h]
\begin{center}
\includegraphics[scale=.45]{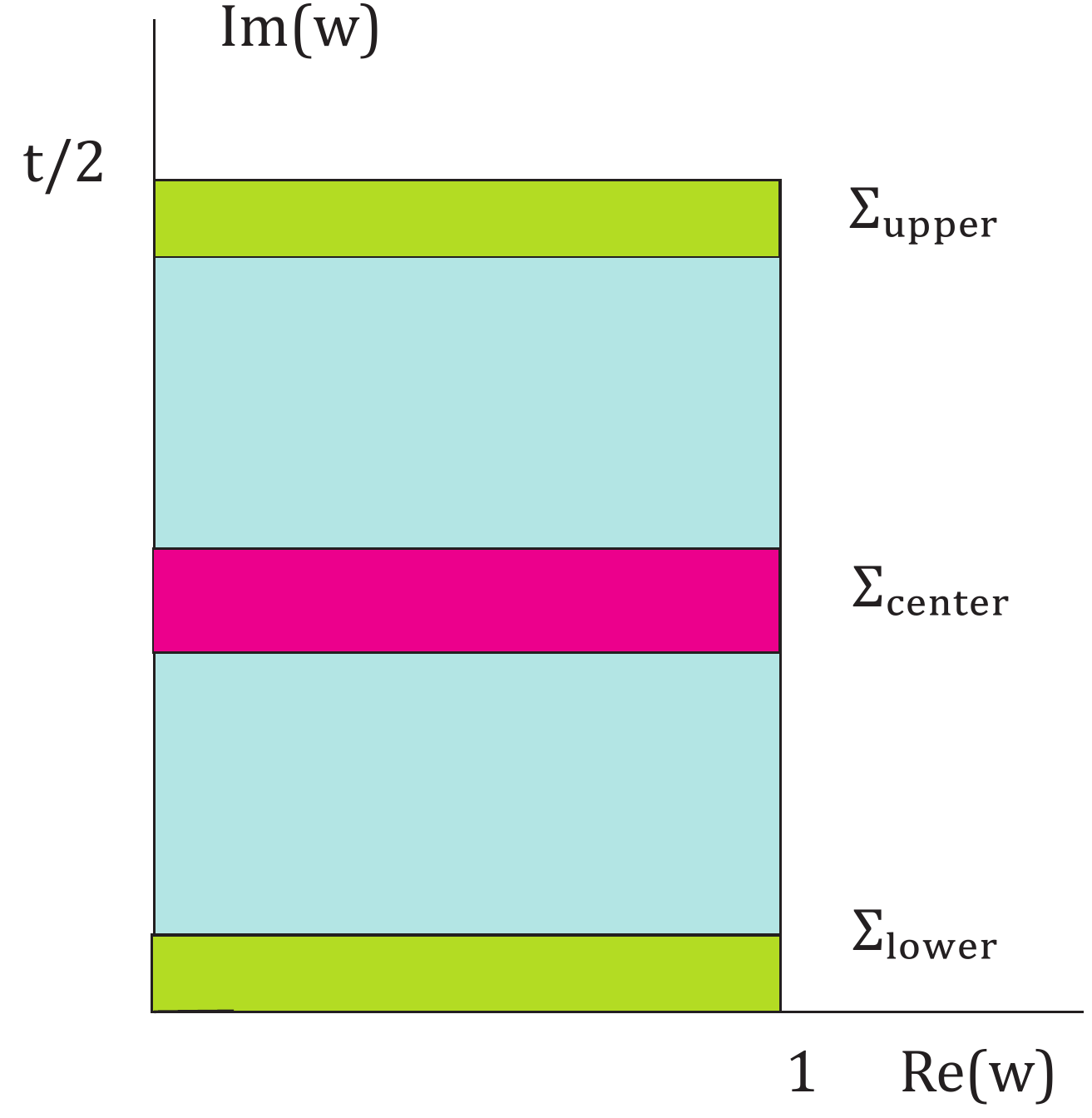} 
\caption{The regions $\Sigma _{\rm lower}$, $\Sigma _{\rm center}$, and $\Sigma _{\rm upper}$.} 
\label{annfig7}
\end{center}
\end{figure}

The basic meromorphic functions then take the following form,
\bea
\label{5b3}
\zeta _0 ( w- x ) = \pi { 1 + u^2 \over z-u} + { 2 \pi (w -x)\over t}  + \pi u 
\eea
Here, we have not converted the term in $w-x$ because its contribution will 
dominate in the region $\Sigma _{\rm center}$ where $w$ is a suitable coordinate.
The expressions for the basic harmonic and holomorphic
functions are then given by,
\bea
\label{5b4}
H & = &  
- 2 \, \Im   \sum _{n=1} ^N c_n \left ( \pi { 1 + u_n^2 \over z-u_n} + { 2\pi w \over t}  \right ) 
-2 \, \Im  \sum _{n'=1} ^{N'} c'_{n'} \left ( \pi  { 1 + u_{n'}'^2 \over z' + u'_{n'}} + { 2 \pi w' \over t} \right ) 
\\
L^A & = &  L_0^A + \sum _{p=1}^P \ell ^A _p \left ( \pi { 1+v_p^2 \over z- v_p}  + { 2 \pi w \over t} + \pi v_p \right ) 
+ \sum _{p'=1}^{P'} \ell '^A _{p'} \left ( \pi { 1+v_{p'}'^2 \over z' + v'_{p'}}  + { 2 \pi w' \over t} - \pi v'_{p'}  \right ) 
\no 
\eea
For large $t$, the behavior  in the different regions is as follows. 
In $\Sigma _{\rm lower}$ the terms in $1/t$ and in $z'$ may be neglected to
leading approximation.  In particular, we have $z' = i + \cO(e^{-\pi t})$ in this region, 
so that the contributions of the $z'$ terms to $H$ and $L_0^A$ vanish to this order. 
Thus, in the region $\Sigma _{\rm lower}$ we have,
\bea
\label{5b5}
H = - 2\pi \, \Im   \sum _{n=1} ^N c_n  { 1 + u_n^2 \over z-u_n} 
\hskip 1in 
L^A = L_0^A + \pi \sum _{p=1}^P \ell ^A _p  { 1+v_p^2 \over z- v_p}  
\eea
up to corrections suppressed by $e^{-\pi t}$.
Similarly, in the region $\Sigma _{\rm upper}$ we retain only the terms in $z'$.
The corresponding expressions for $H$ and $L^A$ may be read off from (\ref{5b4}).

\sm

It remains to determine the behavior of the solution in the region $\Sigma _{\rm center}$.
It will be convenient to parametrize $w=\alpha + it \beta /2$, so that the region $\Sigma _{\rm center}$
corresponds to $0\leq \a <1$ and $|\beta -1/2| \ll 1$. In this regime, we have $z = i + \cO(e^{-\pi t/2})$
as well as $z' = i + \cO(e^{-\pi t/2})$, so that all terms involving $z$ and $z'$ cancel out of $H$ and $L^A$
in (\ref{5b4}) to this order, and we are left with, 
\bea
\label{5b6}
H & = &  (1-\beta ) \cC + \beta \cC' 
\no \\
L^A & = &  L_0^A -i K_0^A + \left ( \a + { it \over 4} ( 2\beta -1 ) \right ) K^A 
\eea
where we have defined $K^A$ as the monodromy in (\ref{3c2}), and the remaining quantities by, 
\bea
\cC = 2 \pi \sum _{n=1}^N c_n 
\hskip 0.6in 
\cC' = 2 \pi \sum _{n'=1}^{N'} c'_{n'} 
\hskip 0.8in
K_0^A = { \pi \over 2} \left ( \sum _{p=1}^P \ell ^A _p + \sum _{p'=1} ^{P'} \ell '^A _{p'} \right )
\eea
We see that near $\beta =1/2$, $H$ and $L^6$ are approximately non-zero constants, 
while the other $L^A$ are linear in $w-\tau/4$. Thus, the geometry is that of the 
product space,
\bea
AdS_2 \times S^2 \times S^1 \times I_\beta
\eea
where $I_\beta$ is the interval in $\beta$ for the region $\Sigma _{\rm center}$.
Through its dependence on $\alpha$, the functions $L^A$ retain their monodromy,
thus transferring monodromy also to the scalars. This geometry is the realization of the 
funnel, but now supplemented with the necessary monodromy to make funnels 
on the asymptotic disks possible with non-zero charge transfers.

\newpage

\section{Half-BPS string-junction solutions at higher genus}
\setcounter{equation}{0}
\label{sec6}

The set-up for string-junction solutions in Type IIB in the presence of multiple boundary
components was given in \cite{Chiodaroli:2009xh} for the special case of $SO(2,2)$
holonomy and genus 0. In the present section, we shall relax both restrictions, and 
construct string-junction solutions with general $SO(2,m)$ holonomy, with an arbitrary
number of boundary components $\nu$, and arbitrary genus $g$. We begin by reviewing 
some standard results on higher genus bordered surfaces, discussed generally in 
\cite{D'Hoker:1988ta,Fay}.

\subsection{Higher genus set-up with boundary components}

\begin{figure}[htb]
\begin{center}
\includegraphics[scale=.57]{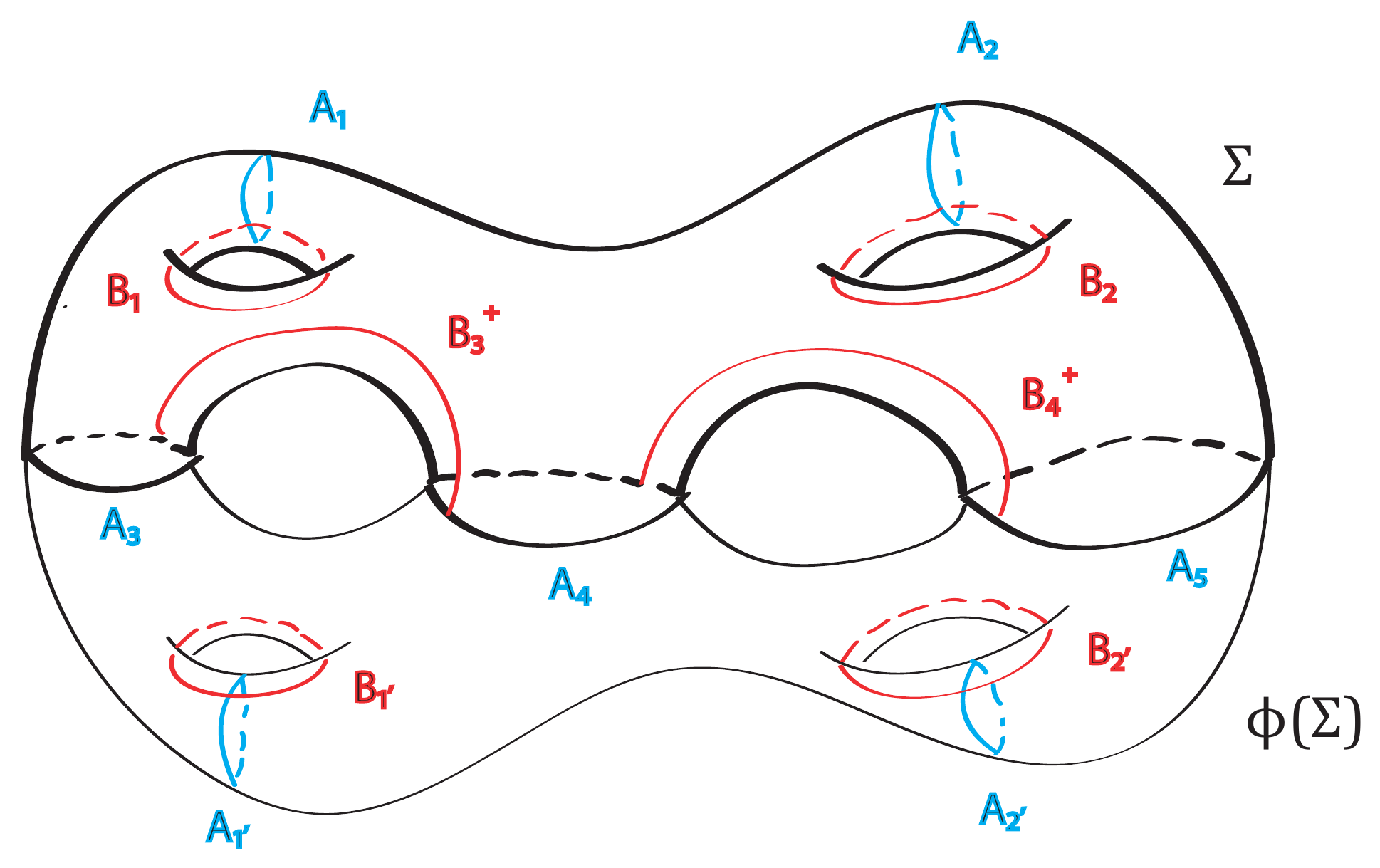} 
\caption{The Riemann surface $\Sigma$ of genus $g=2$ with $\nu=3$ boundary components
$A_3, \, A_4 , \, A_5$, and its double compact Riemann surface  $\hat \Sigma = \Sigma \cup \phi (\Sigma)$ 
of genus $\hat g=6$.}
\label{annfig2}
\end{center}
\end{figure}

Consider an orientable Riemann surface $\Sigma$ of genus $g$ with $\nu$ boundary components.
As usual, we construct functions and forms on $\Sigma$ in terms of  functions and forms on
the double cover Riemann surface $\hat \Sigma$, restricted under the anti-conformal involution $\phi$. 
The boundary $\p \Sigma$ of $\Sigma$ is fixed under $\phi$, so that $\phi (\p \Sigma) = \p \Sigma$,
and we may view $\Sigma$ as the quotient $\Sigma = \hat \Sigma / \phi$. The genus $\hat g$ of 
$\hat \Sigma$ is related to $g$ and $\nu$ by,
\bea\label{genusdoub}
\hat g = 2g + \nu -1
\eea
This construction is depicted schematically in Figure \ref{annfig2} for the case $g=2$ and $\nu=3$. 

It will be useful to sort homology cycles, and their dual holomorphic 1-forms, according to their behavior 
under the involution $\phi$. Our labeling generalizes the one indicated in Figure~\ref{annfig2}, 
\bea
\label{4a1}
\hbox{$2g$ cycles belonging to $\Sigma$~} & \hskip 0.6in & A_I, \, B_I, \hskip 0.5in I=1, \ldots, g
\\
\hbox{$2g$ cycles conjugate under $\phi$~~} && A_{I'}, \, B_{I'}
\no \\
\hbox{$\nu$ boundary cycles for $\p \Sigma$} &&  A_i,  \hskip 0.5in i=g+1, \ldots, g+\nu
\no \\
\hbox{$\nu-1$ conjugate cycles for $\p \Sigma$} &&  B_i,  \hskip 0.5in i=g+1, \ldots, g+\nu-1
\no
\eea
Note that the boundary cycle $A_{g+\nu}$ is homologically dependent upon the other above cycles.
The basis of cycles, and their orientations, may be chosen such that,
\bea
\label{4a2}
A_{I'} = + \phi(A_I) & \hskip 1in & \phi (A_i)  = + A_i
\no \\
B_{I'} = - \phi (B_I) && \phi (B_i) = - B_i
\eea
Throughout,  indices $I$ and $i$ will be running over the ranges defined in (\ref{4a1}).
The corresponding holomorphic differentials will be denoted by $\o_I, \o _{I'}$, and $\o_i$, 
and may be canonically normalized by the following relations, 
\bea
\label{4a3}
\oint _{A_I} \o_J = \oint _{A_{I'}} \o_{J'} = \delta _{IJ} \hskip 1in \oint _{A_i} \o_j = \delta _{ij}
\eea
all other integrals over $A$-cycles being zero. As a result of (\ref{4a2}), we then have the following 
involution relations for the holomorphic differentials,
\bea
\label{4a4}
\phi^* (\o_{I}) & = & \bar \o_{I'}
\no \\
\phi ^* (\o_{I'})  & = &  \bar \o_I 
\no \\
\phi ^* (\o_i)  & = &  \bar \o_i
\eea
where $\phi^*$ is the pull-back of $\phi$ to differential forms. The period matrix of $\hat \Sigma$
is defined by, 
\bea
\label{4a5}
\oint _{B_{\hat I}} \o_{\hat J} = \Omega _{\hat I \hat J}
\eea
where we make use of the composite indices $\hat I = ( I, I',i)$ and $\hat J = (J,  J' ,j)$.
The matrix is symmetric $\Omega _{\hat J \hat I} = \Omega _{\hat I \hat J}$, and has positive definite 
imaginary part $\Im (\Omega) > 0$.
As a result of the involution relations (\ref{4a3}), there exist further relations between the
entries of $\Omega _{\hat I \hat J}$, deduced from the following integral relations, 
\bea
\label{4a6}
\Omega _{i j} = \oint _{B_i} \o_j & = & - \oint  _{\phi (B_i)} \phi ^* (\bar \o _j) 
= - \oint _{B_i} \bar \o_j = -\bar \Omega _{ij}
\no \\
\Omega _{i J'} = \oint _{B_i} \o_{J'} & = & - \oint  _{\phi (B_i)} \phi ^* (\bar \o _J) 
= - \oint _{B_i} \bar \o_J = -\bar \Omega _{iJ}
\no \\
\Omega _{I'J'} = \oint _{B_{I'}} \o_{J'} & = & - \oint  _{\phi (B_I)} \phi ^* (\bar \o _J) 
= - \oint _{B_I} \bar \o_J = -\bar \Omega _{IJ}
\no \\
\Omega _{I'J} = \oint _{B_{I'}} \o_{J} & = & - \oint  _{\phi (B_I)} \phi ^* (\bar \o _{J'}) 
= - \oint _{B_I} \bar \o_{J'} = -\bar \Omega _{IJ'}
\eea
Thus, $\Omega$ takes on the following block form, where the indices run over $\hat I = (I, I', i)$,
\bea
\label{4a7}
\bar \Omega = - \cK \Omega \cK^t 
\hskip 0.8in 
\cK = \left ( \matrix{ 0 & I_g & 0 \cr I_g & 0 & 0 \cr 0 & 0 & I_{\nu-1} \cr } \right )
\eea
where $I_g$ and $I_{\nu-1}$ are the identity matrices respectively in dimensions $g$ and $\nu-1$.

\subsection{Construction of the harmonic function $h_0$}

As is familiar from the case of the annulus, the harmonic functions $H$
and $h^A$ may all be built up out of a single type of elementary harmonic function
$h_0(w|x)$ which has a single simple pole on the boundary at $x \in \p \Sigma$. 
In terms of meromorphic functions $Z_0$ we have,
\bea
\label{4b1}
h_0(w|x)= i Z_0(w|x,q) - i Z_0(\phi(w)|x,q)
\eea
where $Z_0(w|x_0,q)$ is an Abelian integral of the second kind, with a single simple 
pole at~$x$. The auxiliary  point $q$ will drop out 
of  $h_0$ provided $q \in \p \Sigma$.  A convenient building block for Abelian 
integrals of the second kind is the prime form $E(z,w)$ for the surface $\hat \Sigma$,
as well as the holomorphic Abelian 1-forms $\o_{\hat I}$ on $\hat \Sigma$.
In terms of those, we have,
\bea
\label{4b2}
Z_0(w|x,q)= \int _q ^w dy \left ( \p_x \p_y \ln E(x,y) + \sum _{\hat I} \a_{\hat I} (x) \o _{\hat I} (y) \right )
\eea
where the complex 1-forms $\a_{\hat I}(x)$ remain to be determined. To do so, we use the 
conditions:
\begin{enumerate}
\itemsep=-0.05in
\item $h_0$ must be single-valued around the cycles $A_I$, $B_I$, $A_{I'}$, and $B_{I'}$ for $I=1,\ldots, g$;
\item $h_0$ must vanish along the cycles $A_i$ for $i=g+1, \ldots, g+\nu$;
\item $h_0$ must be positive inside $\Sigma$, a condition which will follow from 
the above requirements.
\end{enumerate}

\sm

We begin by implementing condition 1. on the cycles $A_I$ and $A_{I'}$. The first term in (\ref{4b2}) 
has trivial monodromy under all $A$-cycles. Computing the monodromy under general $A_{\hat I}$-cycles, 
the contribution of the second term of Abelian integrals of the first kind gives,   
\bea
\label{4b3}
Z_0 (w+A_{\hat I} |x,q) = Z_0 (w|x,q) + \a_{\hat I}(x)
\eea
Single-valuedness of $h_0$ under the cycles $A_I$ and $A_{I'}$ thus requires $\a_I, \a_{I'} \in \bR$.
Next, we implement condition 1. on cycles $B_I$ and $B_{I'}$. This time both terms in (\ref{4b2}) contribute
to the monodromy, and we use the well-known relation,
\bea
\label{4b4}
\oint _{B_{\hat I}} dy \, \p_x \p_y \ln E(x,y) = 2 \pi i \o_{\hat I} (x)
\eea
We find, 
\bea
\label{4b5}
Z_0 (w+B_{\hat I} |x,q) = Z_0 (w|x,q) + 2 \pi i \o_{\hat I} (x) + \sum _{\hat J} \a_{\hat J} (x) \Omega _{\hat I \hat J}
\eea
Since $\phi^* (\o_{\hat I}) = \bar \o_{\hat I}$, and $\phi (x)=x$,  single-valuedness of
$h_0$ around the cycles $B_I$ and $B_{I'}$ requires the following relations, 
\bea
\label{4b6}
0 & = & 
-2 \pi  \o_{I} (x) -2 \pi  \bar \o_{I} (x) + i \sum _{\hat J} \a_{\hat J} (x) \Omega _{I \hat J} 
 - i \sum _{\hat J} \bar \a_{\hat J} (x) \bar \Omega _{I \hat J} 
\no \\
0 & = & 
-2 \pi  \o_{I'} (x) -2 \pi  \bar \o_{I'} (x) 
+ i \sum _{\hat J} \a_{\hat J} (x) \Omega _{I' \hat J}  - i \sum _{\hat J} \bar \a_{\hat J} (x) \bar \Omega _{I' \hat J}
\eea
Before solving these equations, we first derive the consequences of condition 2.

\sm

To implement condition 2, we make use of the conjugation property of the prime form,
\bea
\label{4b7}
\phi^* E(x,y) = \overline{E(x,y)}
\eea
If $x,y$ belong to the same boundary component, then this equation implies that 
$E(x,y)$ is real. If, however, $x,y$ belong to different boundary components, then 
$E(x,y)$ will not in general be real. We begin by considering the
case where $x $ belongs to a definite boundary component, which we take to be $A_{g+\nu}$.
Since $h_0$ does not depend on $q$, we shall assume that also $q \in A_{g+\nu}$. As a result
of (\ref{4b7}), we see that  $Z_0 (w|x,q)$ is real when $w \in A_{g+\nu}$, so that
\bea
\label{4b8}
h_0(w|x)=0 \hskip 1in \hbox{when} ~ w,x\in A_{g+\nu}
\eea
The same argument could have been made for $w,x$ belonging both
to any given one of the $A_i$-cycles of $\p \Sigma$, but the issue at hand here 
is whether the choice can be made {\sl consistently simultaneously on all the cycles  $A_i$}.
Two different $A_i$ cycles   are separated by a sum of half $B_i$-cycles.
Thus, it suffices to investigate the relation between $h_0$ evaluated on points $w$
and $w'$ which differ by a half-cycle $B_i^+$. To this end, we calculate,
\bea
\label{4b9}
h_0 (w'|x) - h_0(w|x) & = & i \left ( \int _{B_i^+} - \int _{B_i^-} \right )
\bigg ( dy \, \p_y \p_x \ln E(x,y) + \sum _{\hat J} \a_{\hat J} (x) \o_{\hat J} \bigg )
\no \\ & = &
i \oint _{B_i} \bigg ( dy \, \p_y \p_x \ln E(x,y) + \sum _{\hat J} \a_{\hat J} (x) \o_{\hat J} \bigg )
\eea
The integral may be carried out, and the result may be decomposed as follows, 
\bea
\label{4b10}
h_0 (w'|x) - h_0(w|x) = - 2 \pi \o_i (x) + i \sum _{J} \a_{J} (x) \Omega _{i J} + i \sum _{J'} \a_{J'} (x) \Omega _{i J'}
+ i \sum _{j} \a_{j} (x) \Omega _{i j}
\eea
The left side is real by construction, requiring also the right side to be real. Clearly, $\o_i(x)$ is 
real for $x \in \p \Sigma$. Since $\Omega _{ij}$ is purely imaginary, $\a_i$ must be real. 
Thus,  all coefficients $\a_{\hat I}$ must be real. From the second line in (\ref{4a6}),
we have $\Omega _{iJ'} = - \bar \Omega _{iJ}$, from which we see that the  remaining terms
in (\ref{4b10}) will be real provided we require also $\a_{J'} = \a_J$. Setting  $h_0 (w'|x) = h_0(w|x)$  for all 
$B_i$-cycles will precisely enforce all the required Dirichlet boundary conditions. 
Assembling these conditions with those derived in (\ref{4b6}), and using the reality 
of $\a_{\hat I}$,  then gives the following set of conditions, 
\bea
\label{4b12}
\sum _{\hat J} \Im \left ( \Omega _{\hat I \hat J} \right ) \a_{\hat J} (x) 
= - \pi  \o_{\hat I} (x) - \pi  \bar \o_{\hat I} (x)  
\eea
The matrix $\Im (\Omega)$ is  invertible, and may be used to solve uniquely for $\a_{\hat I}$. 
To complete the determination of the harmonic function $h_0$, it remains to 
establish its positivity.

\subsection{Positivity of $h_0(w|x)$}

For the construction of $H$ and $h^A$ to work, the harmonic function $h_0(w|x)$ must be positive.
Note that the electrostatics interpretation of $h_0$ is that of the electric potential produced
by a point-like electric dipole with unit strength, placed at the boundary point $x$,
and with positive polarity facing inward into $\Sigma$, while maintaining the boundary
grounded. It is intuitively clear that the potential inside $\Sigma$ must then always 
be positive. A mathematical proof of the same fact may be given by using the 
{\sl Minimum-Maximum Theorem} for harmonic functions.  It states 
that if $h_0$ is harmonic, then on every compact subspace $K$ (of whatever 
manifold on which the harmonic function lives), the function $h_0$ will attain its 
minimum and its maximum on the boundary $\p K$ of $K$.  

\sm
To apply the theorem here, we consider any subspace 
$\Sigma _\ep$ of $\Sigma$ which avoids the singularity, and on which $h_0$
is a smooth harmonic function. For a general choice of $\Sigma _\ep$, 
the values of $h_0$ along $\p \Sigma _\ep$ are unknown to us.
A useful choice for $\Sigma _\ep$ is obtained by  removing from  $\Sigma$
a small half-disk of radius $\ep$ around the pole at $x$. The boundary of $\Sigma_\ep$
now consists of the boundary of $\Sigma$ with a small half-circle around the pole
replacing the line segment through the pole. On the boundary of $\Sigma$ part,
$h_0$ vanishes, while on the half-circle part we can compute the value of $h_0$
approximately for sufficiently small $\ep$ since the behavior there is dominated 
by the pole. But near the pole $h_0$ is positive; hence $h_0$ must be positive 
throughout $\Sigma$.

\subsection{Construction of  $H$, $h^A$, and $L^A$}

Transcribing the Type IIB solution with holonomy $SO(2,2)$ and genus 0  to the case
of the 6-dimensional supergravity, extending to full $SO(2,m)$ holonomy, as
well as generalizing to arbitrary genus $g$, 
we see that the structure in light-cone variables is very similar to the one derived
for the annulus. Indeed, the fundamental functions are given by,
\bea
\label{4d1}
H(w) & = &  
\sum _{n=1}^N  c_n \, h_0 (w|x_n) 
\no \\
h^A( w) & = &
- \half \sum _{p=1}^P  \ell ^A _p \, h_0  (w|y_p)
\no \\
L^A (w) & = & L^A_0 + \sum _{p=1}^P \ell ^A_p Z_0 (w|y_p,q)
\eea
for $A=1,6,7, \ldots, m+5$.
The functions $h_0$ and $Z_0$ were constructed in the preceding sections. The points $x_n$ 
and $y_p$ lie on the boundary $\p \Sigma$, and will in general be distributed across the 
different boundary components. The above notation subsumes
the one used for the annulus, so that the points $x_n$ and $x'_\np$ of the
annulus are collected here into a single notation of points $x_n$, and similarly
for the points $y_p$.  The residues $c_n$ and $\ell ^A _p$ are real, and we have 
$c_n >0$. Finally, the harmonic function $h_0$ may be expressed in terms of 
a meromorphic Abelian integral $Z_0$ of the second kind, as was done in (\ref{4b1}).

\subsection{Single-valuedness of $L^6$}

In the light-cone parametrization of the solutions, it is essential that the light-cone 
function $\lambda ^- = 1/(\sqrt{2} L^6)$, and thus $L^6$ be single-valued on $\Sigma$.
This means absence of monodromies around the cycles $A_I, B_I, A_{I'}, B_{I'}$, as 
well as the boundary cycles  $A_i$. Since the boundary cycle $A_{g+\nu}$ is homologically
a linear combination of $A_i$-cycles, no condition on $A_{g +\nu}$ needs to be imposed separately.
Making use of the monodromy equations (\ref{4b3}) and (\ref{4b5}), as well as 
condition (\ref{4b12}), we have the following monodromies for $Z_0$,
\bea
\label{4e1}
Z_0 (w+A_{\hat I} |x,q) & = & Z_0 (w|x,q) + \a_{\hat I}(x)
\no \\
Z_0 (w+B_{\hat I} |x,q) & = & Z_0 (w|x,q) + \b_{\hat I}(x)
\eea
where $\a_{\hat I}$ and $\beta _{\hat I}$ are real, and $\beta_{\hat I}$ is given by,
\bea
\label{4e2}
\beta _{\hat I} (x) = i \pi \Big ( \o_{\hat I} (x) - \bar \o_{\hat I} (x) \Big )
+ \half \sum _{\hat J}  \left ( \Omega _{\hat I \hat J} + \bar \Omega _{\hat I \hat J} \right ) \a_{\hat J}(x)
\eea
It is immediate from the relations $\a_{I'}=\a_I$ and (\ref{4a6})  that we have,
\bea
\label{4e3}
\beta _{I'} (x) = - \beta _I (x) 
\hskip 1in
\beta _i (x) = 0
\eea
The absence of monodromy for $A_I$- and $B_I$-cycles implies the absence of monodromy
for $A_{I'}$- and $B_{I'}$-cycles, since we have $\a_{I'}=\a_I$ and $\beta _{I'}(x) = - \beta _I(x)$.
Together with the absence of monodromy around $A_i$-cycles, these conditions 
restrict the residues $\ell ^6 _p$ as follows, 
\bea
\label{4e4}
\sum _{p=1}^P \ell ^6_p \, \a _I (y_p)= \sum _{p=1}^P \ell ^6_p \, \b _I (y_p)
=\sum _{p=1}^P \ell ^6 _p \, \alpha _i (y_p)=0
\eea
which must hold for all $I=1,\ldots, g$, and $i=g+1, \ldots, g+\nu -1$. 

\sm

As was the case for the annulus, an alternative construction for $L^6$ may be given 
in terms of a multiplicative formula, by expressing the 1-form $\cN(w) = \p_w H/L^6$ in 
terms of  the prime form $E(z,w)$ and the 
holomorphic form $\sigma (z)$. The former has weight $(-1/2,0)$ in $z$ and $w$, 
and the latter has weight $(\hat g/2,0)$ in $z$. Both were introduced by Fay in \cite{Fay}, and reviewed
in \cite{Chiodaroli:2009xh}. Here, we shall only need the following properties. Both
$E(z,w)$ and $\sigma (v)$ are free of monodromy around $A_{\hat I}$-cycles, and have
the following monodromy around $B_{\hat I}$-cycles, 
\bea
\label{4e6}
E(z+B_{\hat I} , w) & = & - E(z,w) \exp \Big \{ - i \pi \Omega _{\hat I \hat I} - 2 \pi i \f _{\hat I} (z)
+ 2 \pi i \f _{\hat I} (w) \Big \}
\no \\
\sigma (z+B_{\hat I} )^2  & = & \sigma (z)^2 \exp \Big \{ 2 \pi i (\hat g-1) \Omega _{\hat I \hat I} 
- 4 \pi i  \Delta _{\hat I} + 4 \pi i (\hat g -1) \f _{\hat I} (z) \Big \}
\eea
where $\f _{\hat I}: \hat \Sigma \to \bC^{\hat g} /\Lambda$ is the Abel map, 
and $\Delta $ the Riemann vector, defined by,
\bea
\f_{\hat I} (z) = \int _{z_0} ^z \o_{\hat I} 
\hskip 1in 
\Delta _{\hat I} = \half - \half \Omega _{\hat I \hat I} + \sum _{\hat J \not= \hat I} \oint _{A_{\hat J}} \o_{\hat J} \, \f _{\hat I} 
\eea
The lattice $\Lambda$ is generated by $(I, \Omega)$, The formula for $L^6$ is then built up as follows,
\bea
\label{4e7}
\cN (w) = \cN_0 e^{ -2 \pi i \theta ^t \f (w)} {\prod _{p=1}^P E(w,y_p) 
\over \prod _{n=1}^N E(w,x_n)^2} \sigma (w)^2
\eea
It is readily verified that $\cN$, thus constructed, is automatically a form of weight $(1,0)$ in $w$,
and has the correct poles and zeros. Absence of monodromy around $A$-cycles requires
that all the entries of the column vector $\theta$ be integers. As for the annulus, 
we shall assume that $\cN$  is single-valued on $\hat \Sigma$. The Riemann-Roch theorem
then implies the following relation between the number of poles, zeros, and the genus,
\bea
\label{4e8}
P-2N= 2 \hat g -2
\eea
Absence of monodromy of $\cN$ then reduces to the conditions,
\bea
\label{4e9}
\sum _{p=1}^P \f_{\hat I} (y_p) - 2 \sum _{n=1}^N \f_{\hat I} (x_n) - 2 \Delta _{\hat I} 
=  \theta '_{\hat I} + \sum _{\hat I} \Omega _{\hat I \hat J} \theta _{\hat J} 
\eea
where $\theta _{\hat I}$ and $\theta ' _{\hat I}$ are integers. Since the right side belongs to $\Lambda$,
this condition is just the usual canonical divisor condition for meromorphic 1-forms.

\subsection{Reality of $L^6$ on all boundary components}

By construction, $\cN(w)$ is single-valued on $\Sigma$, including around the homology
cycles $A_I, B_I$ and the boundary cycles $A_i$. It remains to ensure that $L^6 (w)$,
as defined by $L^6(w)= \p_w H /\cN(w)$,  is real on all boundary components. This
in turn requires that $\cN(w)$ should be purely  imaginary on all boundary components.
Since $\cN(w)$ is defined only up to a multiplicative complex constant $\cN_0$, it will suffice to
show that, 
\begin{enumerate}
\itemsep=-0.05 in
\item The phase of $\cN(w)$ is constant (independent of $w$) on each boundary component.
\item The constant phases of $\cN(w)$ on different boundary components are the same.
\end{enumerate}
It will then follow that the phase of the overall multiplicative constant $\cN_0$ may be chosen 
so that $\cN(w)$ is purely imaginary on all boundary components.

\subsubsection{The phases of the prime form on boundary points}

We begin by calculating the phases of the prime form $E(u,v)$, evaluated between  
points $u,v$ on the boundary of $\Sigma$. When $u,v$ belong to the same 
boundary component $E(u,v)$ is real. Next, we assume that $u \in A_i$ and $v \in A_{i+1}$.
One may represent the point $v$ as the image under a half $B_i$-period of a point $v'$ on $A_i$,
namely $v = v' + B_i^+$ with $B^+_i \subset \Sigma$. We then have,
\bea
\label{9a1}
{E(u,v) \over E(u,v)^*} = {E(u,v'+B_i^+) \over E(u,v' +B_i^-)} = {E(u,v ) \over E(u,v -B_i)}
\eea
Using formula (\ref{4e6}) for the $B_i$-cycle shift of the prime form, we find, 
\bea
\label{9a3}
{E(u,v) \over E(u,v)^*} = - \exp \left  \{  i \pi \Omega _{ii} - 2 \pi i \int ^v _ u \o _i \right \}
\eea
where the integral is taking along a curve inside $\Sigma$ which is homologous to a sum of half-periods $B_i^+$. 
Despite its appearance, the argument of the exponential in (\ref{9a3}) is purely imaginary. To see this,
we take its real part, which is found to be,
\bea
\label{9a4}
i \pi \Omega _{ii} - \pi i \int ^v _ u \o_i + \pi i \overline{\int ^v _ u \o_i } 
= i \pi \Omega _{ii} - \pi i \oint _{B_i} \o_i =0
\eea
Thus, we may write instead of (\ref{9a3}) a simpler formula in which this property is manifest, 
\bea
\label{9a5}
{E(u,v) \over E(u,v)^*} = - \exp \left  \{   - 2 \pi i \, \Re \int ^v _ u \o _i \right \}
\eea
Recall that this result was derived when $u \in A_i$ and $v \in A_{i+1}$.

\sm

Next, we proceed to the general case with $u \in A_i$ and $v \in A_j$. In view of the property 
$E(u,v)=-E(v,u)$, we may choose $g+1 \leq i < j \leq g+\nu$ without loss of generality.
To compute the phase of $E(u,v)$, we proceed analogously to the earlier case, and we have,
\bea
\label{9a6}
{E(u,v) \over E(u,v)^*} =  {E(u,v) \over E(u,v - \sum _k c_{i,j}^k B_k)}
\eea
where we have defined the following symbol, 
\bea
\label{9a7}
c_{i,j}^k = \left \{ 
\matrix{ +1 & \hbox{if} & i \leq k < j \cr -1 & \hbox{if} & j \leq k < i \cr 0 && \hbox{otherwise} \cr}
\right .
\eea
The repeated index $k$ must be summed over the range $g+1\leq k \leq g+\nu-1$.
Immediate, and useful, properties of the symbol $c_{i,j}^k$ are the following relations, 
\bea
\label{9a7a}
c^k_{j,i} +  c^k _{i,j} & = & 0
\no \\
c_{i+1, j} ^k - c_{i,j}^k & = & \delta _{ik}
\eea
To compute the denominator of (\ref{9a6}), we iterate $j-i$ times the formula (\ref{9a1}), and obtain 
the following generalization of (\ref{9a5}), 
\bea
\label{9a8}
{E(u,v) \over E(u,v)^*} 
= (-)^{i-j} \exp \left  \{  - 2 \pi i \, \sum _k c_{i,j} ^k \Big ( \Re \, \f _k (v) - \Re \, \f_k (u) \right ) \Big \}
\eea
The definition of the symbol $c_{i,j}^k$ was adopted in (\ref{9a7}) so that formula (\ref{9a8}) 
holds for $u \in A_i$ and $v \in A_j$, {\sl without any restriction on the ordering of the indices $i,j$}.

\subsubsection{The phase of $\sigma$ on boundary points}

We make use of the following representation of  $\sigma$ in terms of $\tet$-functions \cite{D'Hoker:1988ta,Fay}, 
\bea
\label{9b1}
{\sigma (u) \over \sigma (u_0)} = { \tet (\f(u) -\sum _\a \f(q_\a) + \Delta ) 
\over \tet (\f(u_0) -\sum _\a \f(q_\a) + \Delta ) } \prod _{\a=1} ^{\hat g} { E(u_0 , q_\a) \over E(u, q_\a) }
\eea
The points $u_0$ and $q_\a$ are arbitrary in the sense that $\sigma (u)$ is independent of their choice.
For simplicity, we will choose the points $u_0$, $q_\a$, as well as the base reference point $z_0$ 
for the Abel map and the Riemann vector, to all belong to the same boundary component, 
which we take to be $A_{g+1}$. It then follows that $\tet (u_0 -\sum _\a q_\a + \Delta )$ and $E(u_0, q_\a)$
are all real. 

\sm

Using the same manipulations that we undertook for the prime form, we now assume that
$u$ belongs to a specific boundary component $A_i$, so that, 
\bea
\label{9b2}
{ \tet (\f(u) -\sum _\a \f(q_\a) + \Delta )  \over \tet (\f(u) -\sum _\a \f(q_\a) + \Delta ) ^*}
= 
{ \tet (\f(u) -\sum _\a \f(q_\a) + \Delta )  \over \tet (\f(u) - c^k _{1,i} B_k  -\sum _\a \f(q_\a) + \Delta ) }
\eea
Using the shift relation for $\tet$-functions, we find, 
\bea
\label{9b3}
{ \tet (\f(u) -\sum _\a \f(q_\a) + \Delta )  \over \tet (\f(u) -\sum _\a \f(q_\a) + \Delta )^* }
=
\exp \left \{ -2 \pi i \sum _k c^k _{g+1,i} \Re \, \left ( \f _k (u) - \sum _\a \f_k (q_\a) + \Delta _k \right ) \right \}
\quad
\eea
Putting all together, we have,
\bea
\label{9b4}
\ln  { \sigma (u)  \over \sigma (u)^*  } 
= \ln { \sigma (u_0)  \over \sigma (u_0)^*  } 
+ 2 \pi i \sum _k c^k _{g+1,i} \bigg (  (\hat g -1) \Re \, \f _k (u) - \Re (\Delta _k) \bigg )
\eea
Note that $q_\a$ does not enter, in accord with the 
fact that $\sigma (u)$ is independent of $q_\a$.

\subsubsection{The phase of $\cN(u)$}

To work out the phase of $\cN (u)$, we label the poles $x_n$ and zeros $y_p$ of $\cN(u)$ in 
a manner that renders explicit to which boundary component each one belongs,
\bea
\label{9c1}
x _n ~ \to ~ x_{j, n_j} & \hskip 1in & j=g+1, \ldots, g+\nu \qquad n_j =1 , \ldots, N_j
\no \\
y_p ~ \to ~ y_{j, p_j} &&  j=g+1, \ldots, g+\nu \qquad p_j =1 , \ldots, P_j
\eea
The $u$-dependence of the phase may then be easily collected from the phases of the prime 
form and of $\sigma$, and we find for $u \in A_i$, and $u_0$ a fixed point in $A_1$,
\bea
\label{9c2}
\ln  { \cN (u) \over \cN (u)^*} 
=
2 \pi i  \sum _k \cL ^k_i \, \Re \, \f _k (u) + 2 \pi i \cM_i + 2 \ln { \sigma (u_0) \over \sigma (u_0)^*}
\eea
where the coefficients $\cL^k _i$ and $\cM_i$ are given by, 
\bea
\label{9c3}
\cL ^k _i & = & -2 \theta ^k + \sum _{j=g+1}^{g +\nu} c^k_{i,j} \left ( P_j
-2N_j \right )  + 2  (\hat g -1) c^k_{g+1,i} 
\\
\cM_i & = & - \sum _j \sum _k c^k_{i,j} \left ( \sum _{p_j=1} ^{P_j}   \Re \, \f _k (y_{j,n_j} ) 
- 2 \sum _{n_j=1}^{N_j}   \Re \, \f _k (x_{j,n_j}) \right ) - 2 \sum _k c^k _{g+1,i} \Re \, \Delta _k
\no 
\eea
All dependence on $u$ on a given boundary component $A_i$ is contained in the 
coefficients $\cL^k_i$, while the $u$-independent part of the phases is contained in $\cM_i$.

\sm

We begin by analyzing the conditions $\cL^k_i=0$. As they stand in (\ref{9c3}), 
the coefficients $\cL ^k_i$ appear to depend on $i$. Evaluating the difference
for two successive values of $i$ using $c_{g+1,i}^k=-c^k_{i,g+1}$,
the second formula in (\ref{9a7a}), and the independence of $\theta ^k$ on $i$,
we find,
\bea
\label{9c4}
\cL ^k _{i+1} - \cL ^k _i = 
\delta _{ik} \sum _{j=g+1}^{g +\nu}  \left ( P_j  - 2 N_j   \right ) - 2 (\hat g -1) \delta _{ik}
= \delta _{ik} ( P-2N-2\hat g +2) 
\eea
which vanishes in view of (\ref{4e8}). As a result, we may set $i=g+1$, 
and  it is consistent to choose the vector $\theta$ as follows, 
\bea
\label{9c6}
2\theta ^k = \sum _{j=k+1}^{g +\nu}  \left ( P_j  -  2N_j    \right ) 
\eea
Each component $\theta ^k$ must be an integer. Enforcing this condition for all $g+1 \leq k \leq g+\nu -1$,
starting at the highest value of $k$ implies that we must have, 
\bea
\label{9c7}
P_j  \, \in \, 2 \bZ \hskip 1in g+1 \leq j \leq g+\nu -1
\eea
It follows from the relation (\ref{4e8}) that we must also have $P_{g+\nu} \in 2 \bZ$.

\sm

Next, we compute the differences of successive $\cM_i^k$, again using (\ref{9a7a}), and we find, 
\bea
\cM_{i+1} - \cM_i & = & 
 - \sum _j  \left ( \sum _{p_j=1} ^{P_j}   \Re \, \f _i (y_{j,n_j} ) 
- 2 \sum _{n_j=1}^{N_j}   \Re \, \f _i (x_{j,n_j}) \right ) + 2 \, \Re \, \Delta _i
\no \\
& = & 
 - \sum _{p=1} ^P    \Re \, \f _i (y_p ) 
+ 2 \sum _{n=1}^N   \Re \, \f _i (x_n)  + 2 \, \Re \, \Delta _i 
\eea
To analyze this equation, we take the real part of (\ref{4e9}).
The term in $\theta _j$ drops out in view of the property $\Re (\Omega _{ij})=0$.
Using also the property $\Re ( \Omega _{iJ'}) = - \Re (\Omega _{iJ})$, we find, 
\bea
\cM_{i+1} - \cM_i = - \theta _i' - \sum _J \Re (\Omega _{iJ}) (\theta _J - \theta _{J'})
\eea
Since the $\theta _i'$ are integers, their contribution is immaterial. For
$\Re (\Omega _{iJ}) \not= 0$, we must have 
\bea
\theta _{J'} = \theta _J
\eea
Under these conditions, the phase $\cM_i$ is independent of $i$, and the overall constant
phases of all boundaries are the same, and may be absorbed into an overall constant factor.

\subsection{Regularity of solutions}

The proof of regularity parallels the proof for the annulus case.
Having constructed a single-valued $L^6$,  the problem of regularity is reduced to
the positivity of $h^A h_A$. But this is easily investigated, and we have,
\bea
\label{4g1}
h^A h_A = { 1 \over 4} \sum _{p, q=1}^P \ell ^A _ p \ell _{qA} h_0(w|y_p) h_0(w|y_q)
\eea
The absence of second order poles in $\lambda ^2$ in the light-cone 
parametrization requires that 
\bea
\label{4g2}
\ell ^A _p \ell _{pA} =(\ell^1_p)^2 - \sum _{R=6}^{m+5} (\ell ^R_p)^2 =0
\eea
for all $p$. Thus, the diagonal terms in the $p,q$ sum vanish, and we are 
left with 
\bea
\label{4g3}
h^A h_A = { 1 \over 4} \sum _{p \not= q} \Big (
|\vec{\ell}_p | \, | \vec{\ell}_q | - \vec{\ell}_p \cdot \vec{\ell}_q \Big )
h_0(w|y_p) h_0(w|y_q)
\eea
where $\vec{\ell}_p = ( \ell ^6_p, \ell ^7_p, \ldots, \ell ^{m+5} _p)$.
By the Schwarz inequality, this quantity is positive term by term, which
proves positivity of $h^Ah_A$, and thus regularity.

\subsection{Monodromy of the solutions}

By construction, the harmonic functions $H$ and $h^A=\Im (L^A)$, for $A=1,7,8,\ldots,m+5$ 
have vanishing monodromy around the cycles $A_I, B_I, A_{I'}, B_{I'}$, for $I=1,\ldots, g$, 
and vanish on the boundary cycles $A_i$, for $i=g+1, \ldots, g+\nu$. In addition, the 
meromorphic function $L^6$ is single-valued on $\Sigma$. Therefore, the meromorphic
functions $L^A$, for $A=1,7,8,\ldots,m+5$, are permitted to have real monodromy 
around the cycles $A_I, B_I, A_{I'}, B_{I'}$, and $A_i$. We will obtain proper stringy solutions 
with  monodromy of the scalar fields belonging to the U-duality group $SO(5,m;\bZ)$ 
provided the monodromy of the associated $\lambda ^A$ belongs to $SO(5,m;\bZ)$ as well.
Thus, we must require that
\bea
\lambda^A (w + C) = M^A{}_B \lambda^B (w)
\eea
with $M\in SO(5,m;\bZ)$, for any homology cycle $C \in H_1 (\Sigma, \bZ)$ on $\Sigma$.
This condition parallels condition (\ref{3e3}) for the annulus, where only a single non-trivial
homology generator is present, corresponding here to $g=0$ and $\nu=2$.

\sm

Next, we evaluate the monodromy matrix on our solutions. The homology cycle $C$
may be decomposed as follows,
\bea
C = \sum _{\hat I} \left ( C_a ^{\hat I} A_{\hat I} + C^{\hat I} _b B_{\hat I} \right )
\hskip 1in C_a ^{\hat I}, \, C_b ^{\hat I} \in \bZ
\eea
where the extended index $\hat I$ runs over $\hat I = (I, I', i)$, and we set $C_b ^i=0$
as $B_i$ is a homology generator of $\hat \Sigma$, but not of $\Sigma$ itself.
Using the basic monodromy relations of (\ref{4e1}), 
\bea
\label{4e1a}
Z_0 (w+A_{\hat I} |y_p,q) & = & Z_0 (w|y_p,q) + \a_{\hat I}(y_p)
\no \\
Z_0 (w+B_{\hat I} |y_p,q) & = & Z_0 (w|y_p,q) + \b_{\hat I}(y_p)
\eea
with $\a _{\hat I} (y_p)$ and $\b _{\hat I}(y_p)$ real, 
\bea
L^A(w +C) & = & L^A(w) + K^A (C)
\no \\
K^A(C) & = & \sum _{p=1}^P \ell ^A _p \sum _{\hat I} 
\left ( C_a ^{\hat I} \a_{\hat I} (y_p) + C^{\hat I} _b \b_{\hat I} (y_p) \right )
\eea
with the understanding that $K^6(C)=0$ for all cycles $C$.
Introducing again the light-cone notation of Section \ref{monola}, we see that the 
accompanying monodromy of $\lambda^A$ works out precisely as it did 
in the case of the annulus in equations (\ref{3e4}) and (\ref{3e5}).
Therefore, the requirement that the monodromy of $\lambda^A$ should 
belong to the U-duality group $SO(5,m;\bZ)$ amounts again to the requirement
that $K^A \in \bZ$, or more restrictively that $2 K^A \in \bZ$ with $2 K \cdot K  \in \bZ$.

\subsection{Counting of parameters}
 
Finally, we shall count the number of moduli parameters of the general supergravity solution 
constructed in  Section \ref{sec6} and provide a physical interpretation for  these parameters.
The counting on a Riemann surface with $g$ handles and $\nu$ boundary components is given in Table~3.  
Here  $\hat g=2g+\nu-1$ is the genus of the doubled Riemann surface $\bar \Sigma$ 
of (\ref{genusdoub}). The number of auxiliary poles $P$ is determined from (\ref{4e8})
by $P= 2N+2 \hat g-2$. Constraints on the parameters are imposed by  relations (\ref{4e9}) which imposes $g+\nu$ conditions on the parameters\footnote{Note that the constraints (\ref{4e9}) for $\hat I=I$ and $\hat I=I'$ are complex conjugate and do not have to be counted separately. Furthermore the imaginary parts of the $\hat I=I$ and $\hat I=i$ constraint are given by sums of B-periods and  give integral points in the Jacobian which do not involve any free continuous parameters.} 
The individual contributions to the counting of the parameters are listed in Table 3. The total number of moduli parameters can be expressed as follows:
\be
\label{countgensig}
2N(m+1) + (2g+\nu-2)(m+2)+ (2g+ \nu -1)m + ( g-1)
\ee
For the case of principal interest, where $m=21$, we find $44(N+\nu)+ 89\,g -68$.

\begin{table}[htdp]
\begin{center}
\begin{tabular}{cc}
parameters  & number of moduli\\
$x_n$& $N$\\
$c_n$ &$ N$\\
$y_p$  & $P$\\
$l_p^R$ &$ (m-1)P$\\
$l^A_\infty$ & $m+1$\\
$\Sigma\;$ {\rm moduli}& $ 6g-6+3\nu$ \\
constraint &$-g-\nu $\\
\end{tabular}
\label{ancountgeneral}
\caption{Counting of the number of moduli parameters 
for the case of general Riemann surfaces with $g$ handles and $\nu$ boundary components.}
\end{center}
\end{table}%

The first term in (\ref{countgensig}) counts the number of charges in the $N$ asymptotic 
$AdS_3 \times S^3$ regions together with  the number  of expectation values of the 
non-attracted scalars in the asymptotic regions.  The $(m+2)$ charges are supported by 
self-dual strings and each asymptotic AdS 
 is  identified with the near-horizon 
region of a semi-infinite string. The attractor mechanism does not fix the values of $m$ scalars 
in the $SO(5,m)/SO(5)\times SO(m)$ coset which determine the expectation values of the moduli of the dual 2-dim CFT.  Note that this part of the counting is completely analogous to the counting for the disk given in Section \ref{resoldisk} and the annulus given in Section \ref{countanu}.

Hence, it is plausible that, just as in the disk and annulus case, the supergravity solution  
based on the general Riemann surface $\Sigma$ can be obtained  by taking a decoupling 
limit of  a junction where $N$ semi-infinite strings come together.

\sm

The second term in (\ref{countgensig}) counts the number of non-contractible one cycles  $C$   inside the Riemann 
surface $\Sigma$. It follows that $m+2$ non-vanishing three form charges can be  supported on a ${\cal C}\times S^2$ cycle.  Note that overall charge conservation reduces the number of independent three form charges by one.

\sm

The third term in (\ref{countgensig}) may be identified with the monodromy of $m$ axionic 
scalars on the Riemann surface $\Sigma$.  A non-vanishing monodromy points to the presence 
of a three-brane in six dimensions. As discussed in Section \ref{sec5} for the case of the annulus a 
self-dual string can end  on a three-brane and absorb the self-dual string charge. 
\sm

The fourth term in  (\ref{countgensig}) is  related to the shape moduli of the Riemann surface. 
We  conjecture that the general supergravity solution is dual to a network of self-dual strings 
and three-branes. In the case of  BPS string networks in ten dimensions   the intersection angles of the  strings are determined by the charges. For networks with 
closed loops there are  additional   shape and size moduli  which can be varied 
\cite{Aharony:1997bh,Bergman:1998gs}.  

\sm

While the  counting of parameters described in this section is intriguing and suggestive, 
we do not have an exact match between 
the string/three-brane network in flat space-time  and our supergravity solution. In particular the role and interpretation of 
handles in $\Sigma$ is still an open, but very interesting problem.

\bigskip

\bigskip
 
\noindent{\Large \bf Acknowledgements}

\bigskip

M.G. gratefully acknowledges the hospitality of the Isaac Newton Institute for Mathematical Sciences, 
Cambridge, U.K. where some of the work presented in this paper was performed.
The work of Eric D'Hoker and Michael Gutperle was supported in part by NSF grant PHY-07-57702. 
The work of Marco Chiodaroli  was supported in part by NSF grant PHY-08-55356.

\newpage

\appendix

\section{Calculation of the Annulus Boundary Entropy}
\setcounter{equation}{0}
\label{seca1}

In this appendix we shall provide the details of the calculation of the boundary entropy 
introduced in Section \ref{boundcalc} for a BCFT with a single asymptotic $AdS_3 \times S^3$
region, with $N=1,\, P=2$ and $N'=P'=0$, in an expansion in which powers of $e^{-\pi t}$ are omitted.

\subsection{Dependence  of $S_A$ on the charges}\label{chargedep}

The formula for the boundary entropy $S_A$ on the annulus simplifies upon
making the choice $x_1=0$, so that we have $y_1=-y_2=y$ and,
\bea
\label{sbounda}
S_A = { 8 \pi \over G_N}  \Big (
|\vec{\ell}_1 | \, | \vec{\ell}_2 | - \vec{\ell}_1 \cdot \vec{\ell}_2\Big )
\int _{\Sigma _\ep} 
d^2w  \, |\cN (w) |^2 h_0(w-y) h_0(w+y)
\eea
where the functions $\cN(w)$ and $h_0(w\pm y)$ are given by, 
\bea
\label{4k2}
\cN (w) & = & \cN_0 { \tet _1 (w-y) \theta _1 (w+y) \over \tet _1 (w)^2}
\no \\
h_0(w \pm y) & = & - \Im \left ( { \p _w \tet _1(w \pm y ) \over \tet _1 (w \pm y)} + { 2 \pi i w \over \tau} \right )
\eea
and we write $\tet _1 (w)$ for $\tet _1 (w|\tau)$. The function $\p_w H$ is purely imaginary
on the boundaries $w=0, \tau/2$, while $L^6$ needs to be real there. Thus, $\cN(w)$ 
must be imaginary on the boundaries as well. Together with the condition $| \cN_0|=1$, 
this will be achieved by setting,
\bea
\cN_0 = -i
\eea 
The choice $\cN_0=+i$ would lead to the same physical quantities.

\sm

The divergent contribution as $\ep \to 0$ arises from the integration over the region near $w=0$
(and its translational copy at $w=1$), where we have the following asymptotic behaviors, 
\bea
\label{app1}
\cN (w)  \sim {\hat \cN _0 \over w^2 } \hskip 0.6in & \hskip 1in &
\hat \cN_0 = i \, { \tet _1 (y|\tau)^2 \over \tet _1 '(0|\tau)^2}
\no \\
h_0(w+y) \sim - \Im (w)  \hat{h}_0(y) && \hat h_0 (y) =  \p_y^2 \ln \tet _1 (y|\tau) + { 2 \pi i \over \tau}
\eea
The divergent part, $S_{\rm div} $ of the entropy integral evaluates to, 
\bea
S_{\rm div}= { 4 \pi^2  \over G_N}   \Big ( |\vec{\ell}_1 | \, | \vec{\ell}_2 | - \vec{\ell}_1 \cdot \vec{\ell}_2\Big ) 
| \hat \cN_0 |^2 \hat{h}_0 (y)^2 \,  \ln {1 \over \ep}
\eea
The divergent term should be proportional to the central charge which in turn is proportional 
to $Q \cdot Q$. The charge (of the single asymptotic $AdS_3 \times S^3$ region) is  given by 
the general formula (\ref{asycha}). We can use the doubling to map the complex conjugate 
part into the lower half-plane and pick up a complete closed contour  around $w=x_1=0$,
\bea
Q^A=  \sqrt{2} \pi \oint_{x_1} dw \partial_w H \lambda^A 
\eea
Performing the contraction of the indices $A$ gives  the invariant square charge $Q\cdot  Q$, 
\bea
Q \cdot Q = 2 \pi^2   \oint_{x_1} dw_1 \; \partial_{w_1} H(w_1)  
\oint_{x_1} dw_2 \; \partial_{w_2} H(w_2) \lambda (w_1) \cdot \lambda (w_2) 
\eea
Using the formula, 
\bea
\lambda (w_1) \cdot \lambda (w_2)-2
= -{1\over {L^6(w_1)L^6(w_2)}} \left (L^A(w_1)-L^A(w_2) \right ) \left ( L_A(w_1)-L_A (w_2) \right )
\eea
where the sum on the right extends over the indices $A=1,6,7, \ldots, m+5$,
we can express the invariant square charge as follows, 
\bea
Q \cdot Q
=-2 \pi^2  \oint_{x_1} dw_1 \;  {\cN(w_1)}  \oint_{x_1} dw_2 \; {\cN (w_2)} 
\left (L^A(w_1)-L^A(w_2) \right ) \Big ( L_A(w_1)-L_A (w_2) \Big )  
\eea
Using the expansions near $w_1=0$ and $w_2=0$ from (\ref{app1}) for $\cN$, 
together with the small $w$ expansion of $\zeta _0$,
\bea
\zeta _0 (w+y) \approx w \hat h_0(y)
\eea
we compute, 
\be\label{a11}
Q \cdot Q= - 32 \pi^4  \Big( |\vec{\ell}_1 | \, | \vec{\ell}_2 | - \vec{\ell}_1 \cdot \vec{\ell}_2\Big )    
\hat \cN _0 ^2    \hat{h}_0(y)^2
\ee
Using the fact that $\hat \cN_0^2 = - | \hat \cN_0|^2$, we find the following formula for the 
entanglement entropy $S_A$ and the boundary entropy $S_{{\rm \scriptstyle bcft}}$,
\be
S_A= {Q \cdot  Q \over 8 \pi^2 G_N}  \ln {1 \over \epsilon} + S_{{\rm \scriptstyle bcft}}
\ee 

\subsection{Asymptotic behavior of the entropy integrals}
\label{asyment}

\begin{figure}[htb]
\begin{center}
\includegraphics[width=2in]{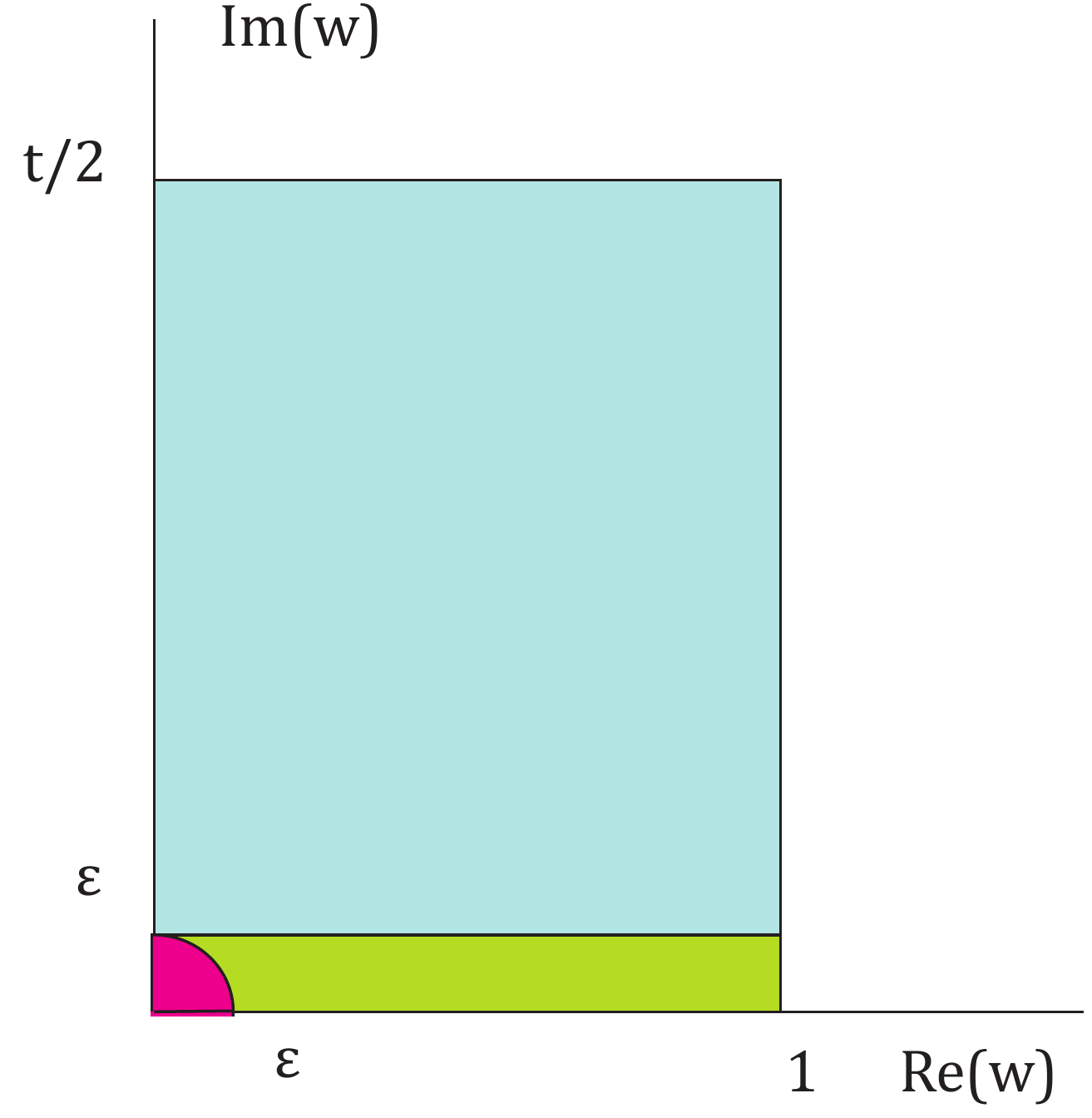} 
\caption{Regularizations $|w|>\ep$ and $\Im (w) > \ep$ inside the fundamental domain.
The domain $\Sigma _\ep$ consists of the union of the blue region and the green region
(minus a corresponding quarter circle around the point $w=1$); the domain 
$\Sigma '_\ep$ consists only of the blue region.}
\label{annfig6}
\end{center}
\end{figure}

The $N=1$, $P=2$ configuration involves the following entropy integral,
\bea
\cJ_\ep (t,y) =  \int _{\Sigma _\ep} d^2w | \cN(w) | ^2   h_0(w-y) h_0(w+y)
\eea
where the ingredients are defined in (\ref{4k2}) with $\cN_0 =-i$.
The shifts by $\pm y$ cancel in the second term 
on the second line above upon taking the imaginary part in view of the fact that $\tau=it$ is purely imaginary.
The integration region is defined as follows, 
\bea
\label{4k1}
\Sigma _\ep & = & \left \{ w = \a + i \beta {t \over 2} ~ 
\hbox{with}  ~~ 0 \leq \a, \b \leq 1 ~~ \hbox{and} ~~ \ep < |w|, ~~ \ep < |1-w| \right \}
\eea
with quarter disks of size $\ep$ removed around the points $w=0,1$, 
as is represented in Figure~\ref{annfig6}. In the figure, we also represent the integration region
$\Sigma _\ep '$, defined by 
\bea
\label{4k3}
\Sigma _\ep ' & = & \left \{ w = \a + i \beta {t \over 2} ~~ 
\hbox{with}  ~~ 0 \leq \a \leq 1 ~~ \hbox{and} ~~ \ep_0 < \b < 1 \right \}
\eea
which will be much easier to use in the actual calculations. Here, we set $ t \ep_0 =2 \ep$.

\subsubsection{Description of the expansion used}

We shall obtain the asymptotic behavior as $t\to \infty $ with $\tau = i t$. The integration region in $w$
will tends to $\infty$, and it is preferable to parametrize the integrand with a fixed range, as
$\a, \b$ do in (\ref{4a1}). We need an expansion for large $\tau$ which is uniform throughout
the integration region. Fortunately, for elliptic functions, this is easy to obtain using
their product formula, 
\bea
\tet _1 (w) =  q_0 \sin \pi w  \prod _{n=1} ^\infty \left ( 1 - 2 q^{2n} \cos (2 \pi w) + q^{4n} \right )
\eea
with $q=e^{i \pi \tau} = e^{-\pi t}$, and $q_0$ an overall $t$-dependent factor which will cancel out
of all our calculations. The behavior of the middle terms in the product, in terms of the fixed range
parametrization of (\ref{4a1}),  are as follows,
\bea
2 q^{2n} \cos (2 \pi w) \sim e^{ \pi i \tau ( 2n -\b ) - 2 \pi i \a}
\eea
Since $0 \leq \beta \leq 1$, all such terms are uniformly suppressed in $\Sigma _\ep$, and
will be omitted. Thus, we are left with the uniform approximation,
\bea
\tet_1(w) \sim q_0 \, \sin \pi w + \cO (q)
\eea
the higher order terms being suppressed by at least one power of $q$.
The corresponding functions of (\ref{4a2}) then take the form, 
\bea
\cN(w)  & = & -i  \, {\sin \pi (w-y) \sin \pi (w+y) \over (\sin \pi w)^2} + \cO (q)
\no \\
h(w\pm y) & = & - { \pi i \over 2} \, { \sin \pi (w-\bar w) \over | \sin \pi (w\pm y) |^2} - { 2 \pi \over t} \Im (w) + \cO(q)
\eea
Thus, the integral $\cJ_\ep$ becomes,
\bea
\cJ_\ep (t,y) & = &
{\pi^2 \over 4} \int _{\Sigma _\ep} {d^2w \over |\sin \pi w|^4} 
\left (  i  \sin \pi (w-\bar w) +{ 4 \over t} \Im (w) |\sin \pi (w+y)|^2 \right )
\\ && \hskip 1in \times 
\left ( i  \sin \pi (w-\bar w) +{ 4 \over t} \Im (w) |\sin \pi (w-y)|^2 \right ) + \cO(q)
\no
\eea

\subsubsection{Changing regulator}

It will turn out that the structure of the integrand, in terms of periodic functions,
is not naturally regularized by the quarter disk cutoffs employed in the definition 
of $\Sigma _\ep$. Instead, the domain $\Sigma _\ep '$ is more natural as the 
integration in $\a$ will then always have the same range. The difference involves
the integral over the region in green in Figure \ref{annfig6}, whose area tends to zero
as $\ep \to 0$, but whose contribution to $\cJ_\ep$ does not.
Thus, we shall decompose the regularized integral as follows,
\bea
\cJ_\ep (t,y) = \cJ _{{\rm reg}} (t,y)
+ \cJ_\ep ^{(0)} (t,y) + { 1 \over t} \cJ_\ep ^{(1)}  (t,y) + { 1 \over t^2} \cJ_\ep ^{(2)}  (t,y) + \cO(q)
\eea
where we have the following expressions, 
\bea
\cJ_\ep ^{(0)} (t,y) & = &  {\pi^2 \over 4} \int _{\Sigma _\ep'} d^2w \, {|\sin \pi (w-\bar w)|^2 \over |\sin \pi w|^4} 
\no \\
\cJ_\ep ^{(1)} (t,y) & = &  \pi^2 \int _{\Sigma _\ep'} d^2w \, { i \sin \pi (w-\bar w) \Im (w) \over |\sin \pi w|^4} 
|\sin \pi (w+y)|^2 + ( y \to -y)
\no \\
\cJ_\ep ^{(2)} (t,y) & = &  4 \pi^2 \int _{\Sigma _\ep'} d^2w \, {  \Im (w)^2 \over |\sin \pi w|^4} 
|\sin \pi (w+y)|^2 |\sin \pi (w-y)|^2 
\eea
and the  term $\cJ_{{\rm reg}}$ accounting for the difference in regulators is given by,
\bea
\cJ_{{\rm reg}} (t,y) & = &
{\pi^2 \over 4} \int _{\left ( \Sigma _\ep - \Sigma _\ep' \right )} {d^2w \over |\sin \pi w|^4} 
\left (  i  \sin \pi (w-\bar w) +{ 4 \over t} \Im (w) |\sin \pi (w+y)|^2 \right )
\\ && \hskip 1in \times 
\left ( i  \sin \pi (w-\bar w) +{ 4 \over t} \Im (w) |\sin \pi (w-y)|^2 \right ) + \cO(q)\no 
\eea
The $y$-dependence of these integrals is easily exposed: $\cJ_\ep ^{(0)} (t,y)$
is simply independent of $y$; while the other two integrals take on the following
$y$-dependence:
\bea
\cJ_\ep ^{(1)} (t,y) & = &  
2 \pi^2 \cos ^2 \pi y \int _{\Sigma _\ep '} d^2w \, { i \sin \pi (w-\bar w) \Im (w) \over |\sin \pi w|^2} \no
 \\ && + 
2 \pi ^2 \sin ^2 \pi y \int _{\Sigma _\ep '} d^2w \, { i \sin \pi (w-\bar w) \Im (w) |\cos \pi w |^2 \over |\sin \pi w|^4} 
\no \\
\cJ_\ep ^{(2)} (t,y) & = &  
4 \pi^2 \cos ^4 \pi y \int _{\Sigma _\ep '} d^2w \,  \Im (w)^2 
\no \\ && +
4 \pi^2 \sin ^4 \pi y \int _{\Sigma _\ep '} d^2w \, {  \Im (w)^2 |\cos \pi w |^4 \over |\sin \pi w|^4} 
\no \\ && -
4 \pi^2 \sin ^2 \pi y \cos ^2 \pi y \int _{\Sigma _\ep '} d^2w \,   \Im (w)^2  
\left ( { \cos ^2 \pi \bar w \over \sin ^2 \pi \bar w} + {\cos ^2 \pi w \over \sin ^2 \pi w} \right )
\eea

\subsubsection{Expansions}

We will evaluate the above integrals by expanding the denominators in a 
uniformly convergent expansion (for $\ep >0$), then carry out the integrals, and re-sum
the results. We shall use the following expansions, 
\bea
\label{exp1}
{ 1 \over \sin \pi w} & = & - 2i \sum _{n=0}^\infty e^{i \pi w (2n+1)}
\no \\
{ 1 \over \sin^2 \pi w} & = & - 4 \sum _{n=1}^\infty n \, e^{2 \pi i n w}
\no \\
{ \cos \pi w \over \sin^2 \pi w} & = & - 2 \sum _{n=0}^\infty (2n+1) e^{i \pi w (2n+1)}
\eea
We shall also use the following resummation formulas in the approximation where 
we neglect corrections suppressed by $e^{-\pi t}$, and contributions which vanish
in the limit where $\ep_0 \to 0$,
\bea
\label{exp2}
\sum _{n=1}^ \infty { 1 \over n} e^{-2 \pi t n \ep_0}  & \approx & - \ln ( 2 \pi t \ep_0 )
\no \\
\sum _{n=1}^ \infty  e^{-2 \pi t n \ep_0}  & \approx & {1 \over 2 \pi t \ep_0} 
\no \\
\sum _{n=1}^ \infty n\,  e^{-2 \pi t n \ep_0}  & \approx & { 1 \over 4 \pi^2 t^2 \ep_0^2}
\eea

\subsubsection{Evaluation of $\cJ_\ep ^{(0)}$}

We begin by writing $ w = \a + i \beta t /2$,
and recast the integral in the following way,
\bea
\cJ_\ep ^{(0)} (t,y) =  2 \pi^2 t \int _0 ^1 d \a \int _{\ep_0} ^1  d \beta \, |\sin i \pi t \b |^2 
\sum _{m,n=1}^\infty mn \, e^{2 \pi i \a (m-n) -  \pi t \b (m+ n)}
\eea
The integral over $\a$ makes the resulting double sum collapse to a single sum,
\bea
\cJ_\ep ^{(0)} (t,y) =  2 \pi^2 t \sum _{n=1}^\infty  n^2 \int _{\ep_0} ^1  
d \beta \, |\sin i \pi t \b |^2 e^{ - 2 \pi t  \beta n}
\eea
Expressing $\sin \pi (w-\bar w)$ in terms of $e^{\pm \pi t \beta}$,
we may rearrange the expansion as follows, 
\bea
\cJ_\ep ^{(0)} (t,y) =  { \pi^2 t \over 2}  + 
{ \pi^2 t \over 2} \sum _{n=1}^\infty   \int _{\ep_0} ^1  d \beta \,  e^{ - 2 \pi t n \beta}
\, \left ( (n-1)^2 + (n+1)^2 - 2 n^2 \right )
\eea
The terms in parentheses simplify to give the value 2, and the remaining integral and sum
are easily carried out, to find, 
\bea
\cJ_\ep ^{(0)} (t,y) =  { \pi^2 t \over 2}  + 
{ \pi \over 2} \ln \left ( 1 - e^{-2 \pi t} \right ) - { \pi \over 2} \ln \left ( 1 - e^{-2 \pi t\ep_0} \right ) 
\eea
To be consistent with the approximation we have made, we must drop the second term 
on the right, and approximate the formula for $ \ep \ll 1$, and we find,
\bea
\cJ_\ep ^{(0)} (t,y) =  { \pi^2 \over 2} t  - { \pi \over 2} \ln \left ( 4 \pi  \ep \right ) 
\eea

\subsubsection{Evaluation of $\cJ_\ep ^{(1)}$}

The evaluation of $\cJ_\ep ^{(1)}$ involves two integrals,
\bea
\cJ_\ep ^{(1)} = \pi^2 \cK_1 \cos ^2 \pi y  + \pi ^2 \cK_2 \sin^2 \pi y 
\eea
where the integrals proper are given by, 
\bea
\cK_1 & = & 2 \int _{\Sigma _\ep '} d^2w \, { i \, \sin \pi (w - \bar w) \Im (w) \over |\sin \pi w|^2}
\no \\
\cK_2 & = & 2 \int _{\Sigma _\ep '} d^2w \, { i \, \sin \pi (w - \bar w) \Im (w) |\cos \pi w |^2 \over |\sin \pi w|^4}
\eea
To compute $\cK_1$, we make use of the first formula in (\ref{exp1}), express $w$
in terms of $\a,\b$, and carry out the integral over $\a$. The result is,
\bea
\cK_1 = t^2 \sum _{n=0} ^\infty \int ^1 _{\ep _0} d \beta \, \beta \left ( 
e^{-2\pi t \beta (n+1)} - e^{-2 \pi t \beta n} \right )
\eea
all terms in the sum, except the second term for $n=0$, cancel one another pairwise, a
so that we are left with,
\bea
\cK_1 = - \half t^2
\eea
To compute $\cK_2$, we make use of the fourth formula in (\ref{exp1}), express $w$
in terms of $\a,\b$, and carry out the integral over $\a$. The result is,
\bea
\cK_2 = t^2 \int ^1 _{\ep _0} d \beta \, \beta 
\sum _{n=0}^\infty (2n+1)^2 \, e^{- \pi t \beta (2n+1)} \left (  e^{- \pi t \beta } - e^{\pi t \beta} \right )
\eea
Combining the terms in the sum by shifting $n\to n-1$ in the first term, and then using $(2n-1)^2-(2n+1)^2=-8n$, 
gives the following simplified sum, 
\bea
\cK_2 = t^2 \int ^1 _{\ep _0} d \beta  \left ( -\b - 8
\sum _{n=1}^\infty \beta n \, e^{- 2 \pi t \beta n}  \right )
\eea
Computing each individual integral, within the approximation of neglecting terms 
suppressed by $e^{-\pi t}$, we find, 
\bea
\int ^1 _{\ep _0} d \beta  \, \beta \, n \, e^{- 2 \pi t \beta n}
\approx 
\left ( { \ep_0 \over 2 \pi t} +{ 1 \over 4 \pi^2 n t^2} \right ) e^{-2\pi t \ep_0}
\eea
Carrying out the sum over $n$ gives, 
\bea
K_2 = - {t^2 \over 2} -{2 \over \pi^2} +{2 \over \pi^2} \ln (2 \pi t \ep_0)
\eea
Putting all together, we find, 
\bea
\cJ^{(1)}_\ep (t,y)= -{\pi^2 t^2 \over 2} - 2 \sin^2 \pi y + 2 \sin^2 \pi y \, \ln (4 \pi  \ep)
\eea

\subsubsection{Evaluation of $\cJ_\ep ^{(2)}$}

The evaluation of $\cJ_\ep ^{(2)}$ involves 3 integrals, 
\bea
\cJ ^{(2)} _\ep = 4 \pi^2 \cos ^4 \pi y \, \cK_3 - 4 \pi^2 \sin^2 \pi y \cos ^2 \pi y \cK_4
+ 4 \pi^2 \sin ^4 \pi y \, \cK_5
\eea
with
\bea
\cK _3 & = & \int _{\Sigma _\ep '} d^2w \, \Im (w)^2
\no \\
\cK _4 & = & \int _{\Sigma _\ep '} d^2w \, \Im (w)^2 \left ( { \cos ^2 \pi w \over \sin ^2 \pi w} 
+ { \cos ^2 \pi \bar w \over \sin ^2 \pi \bar  w} \right ) 
\no \\
\cK _5 & = & \int _{\Sigma _\ep '} d^2w \, \Im (w)^2 \left | { \cos  \pi w \over \sin  \pi w}  \right |^4
\eea
The first integral is trivial to evaluate.
For the second integral, we use the series expansion on the second line of (\ref{exp1}),
and its complex conjugate. Carrying out the $\alpha$-integration, it follows that only the 
$n=0$ in those expansions contributes, so that we get,
\bea
\cK_3 & = & { t^3 \over 24}
\no \\
\cK_4 & = & -2 \cK_3 = - {t^3 \over 12}
\eea
To calculate the integral $\cK_5$  we use again the second line expansion in 
(\ref{exp1}) to proceed,
\bea
\cK_5 = \sum _{m,n=0}^\infty \left ( \delta _{m,0} + 4 m \right ) \left ( \delta _{n,0} + 4 n \right )
\int _{\Sigma _\ep} d^2w \Im (w)^2 \, e^{2 \pi i m w - 2 \pi i n \bar w}
\eea
Carrying out the $\alpha$-integral makes the double sum collapse to only $m=n$ 
contributions in terms on an integral over $\beta$,
\bea
\cK_5 = { t^3 \over 24} + 2 t^3 \sum _{n=1}^\infty \int ^1 _{\ep_0} d \beta \, \beta^2 n^2 \, e^{-2 \pi t \beta n}
\eea
Neglecting again  contributions suppressed by $e^{-\pi t}$, we find,
\bea
\int ^1 _{\ep_0} d \beta \, \beta^2 n^2 \, e^{-2 \pi t \beta n} 
\approx 
\left ( { 1 \over 2 \pi^3 n} + { t \ep _0 \over \pi^2} +{ t^2 \ep_0^2 n \over \pi} \right ) e^{-2 \pi t n \ep_0} 
\eea
Using the resummation formulas of (\ref{exp2}), we can put all parts together gives, 
\bea
\cK_5 = { t^3 \over 24} +{3 \over 4 \pi^3} - { 1 \over 2 \pi^3} \ln (2 \pi t \ep_0)
\eea
The total contribution to $\cJ_\ep ^{(2)}$ is then found to be, 
\bea
\cJ^{(2)} _\ep = { \pi^2 t^3 \over 6} + {3 \over \pi} \sin^4 \pi y -{ 2 \over \pi } \sin^4 \pi y \, \ln (4 \pi  \ep)
\eea

\subsubsection{Evaluation of $\cJ_{{\rm reg}}$}

The regions near $w=0$ and $w=1$ will give equal contributions, which may be approximated 
by the leading term of the integrand near $w=0$, since the divergence is only logarithmic.
That contribution may be approximated by,
\bea
\cJ_{{\rm reg}} = 2   \left ( 1 - { 2 \over \pi t} \sin ^2 \pi y \right )^2 \cI
\hskip 1in 
\cI = \int _{\left ( \Sigma _\ep - \Sigma _\ep' \right )} {d^2w \over  |w|^4} \, \Im (w)^2
\eea
The factor of 2 accounts for the contributions from both $w=0$ and $w=1$. Expressing $w=\ep (x+iy)$,
the integral may be written as follows, 
\bea
\cI
= \int _0 ^ {1/\ep} dx \int _0 ^1 dy { y^2 \over (x^2 + y^2)^2} \theta (x^2+y^2-1)
\eea
In the limit $\ep \to 0$, we may let the upper  integration limit tend to $\infty$. 
Changing variables to polar coordinates, $x = r \cos \psi , ~ y = r \sin \psi$, we obtain, 
\bea
\cI = \int _0 ^{\pi/2} d \psi \int _1 ^{1/\sin \psi} dr \, { \sin ^2 \psi \over r}
= {\pi \over 4} \ln 2 - { \pi \over 8}
\eea
The last integral was obtained using MAPLE.

\subsubsection{Assembling all contributions to $\cJ_\ep$}

Combining the contributions from $\cJ_{{\rm reg}}, \cJ_\ep ^{(0)}, \cJ_\ep ^{(1)}, \cJ_\ep ^{(2)}$, we find,
\bea
\cJ_\ep (t,y) = - { \pi \over 2}   \ln ( 2 \pi \ep ) 
 \left ( 1 - { 2 \over \pi t} \sin^2 \pi y \right )^2 
+{\pi^2 \over 6} t - {\pi \over 4} -{1 \over t} \sin^2 \pi y +{2 \over \pi t^2} \sin^4 \pi y
\eea
Next, we shall express this result with the help of the central charge of the bulk CFTs,
which in turn is proportional to $Q_A Q^A$. We use the relation with the charges, 
\bea
Q \cdot Q = 32 \pi^4 \left ( |\vec{\ell}_1| \, |\vec{\ell}_2| - \vec{\ell}_1 \cdot \vec{\ell}_2 \right ) 
 \left ( 1 - { 2 \over \pi t} \sin^2 \pi y \right )^2
\eea
Thus, we have 
\bea
S_A = - {Q \cdot Q \over 8 \pi^2 G_N}  \ln (2 \pi \ep)  + S_{{\rm \scriptstyle bcft}} 
\eea
where the finite boundary conformal field theory entanglement entropy part $g_{\rm bcft}$ is
given by,
\bea
S_{{\rm \scriptstyle bcft}} 
 = {4 \pi^2 \over G_N} \left ( |\vec{\ell}_1| \, |\vec{\ell}_2| - \vec{\ell}_1 \cdot \vec{\ell}_2 \right ) 
 \left (   {\pi t \over 3} - \half  - { 2 \over \pi t} \sin ^2 \pi y + { 4 \over \pi^2 t^2 } \sin ^4 \pi y \right )
\eea
The dominant term at large $t$ agrees with the funnel.

\end{document}